\documentclass[aps,prd,10pt,groupedaddress,superscriptaddress,showkeys,nofootinbib]{revtex4-2}

\usepackage[driverfallback=dvipdfm]{hyperref}
\usepackage{amsmath,amssymb}
\usepackage{mathtools}
\usepackage{fancyvrb}

\allowdisplaybreaks[1]
\newcommand{\dr}{\mathrm{d}}
\usepackage{color}

\newcommand{\Tr}{\mathrm{Tr}}
\newcommand{\tr}{\mathrm{tr}}
\newcommand{\diag}{\mathrm{diag}}

\renewcommand{\t}{\mathrm{t}}

\renewcommand{\vec}[1]{\boldsymbol{#1}}

\renewcommand{\t}{\mathrm{t}}

\newcommand{\naive}{\mathrm{naive}}
\newcommand{\mirror}{\mathrm{mirror}}

\newcommand{\com}{\mathrm{com}}

\newcommand{\R}{\mathrm{R}}
\renewcommand{\L}{\mathrm{L}}

\newcommand{\U}{\mathrm{U}}
\newcommand{\SU}{\mathrm{SU}}
\newcommand{\SO}{\mathrm{SO}}

\newcommand{\vect}[1]{\boldsymbol{#1}}

\usepackage{ulem}

\renewcommand{\vec}[1]{\boldsymbol{#1}}

\usepackage{bbm}

\usepackage{xcolor}

\usepackage{titlesec}

\begin{document}


\title{Emergent chirality and superfluidity of parity-doubled baryons in neutron stars}


\author{Shigehiro~Yasui}
\email[]{yasuis@keio.jp}
\email[]{s-yasui@nishogakusha-u.ac.jp}
\affiliation{International Institute for Sustainability with Knotted Chiral Meta Matter (SKCM$^2$), Hiroshima University, 1-3-2 Kagamiyama, Higashi-Hiroshima, Hiroshima 739-8511, Japan}
\affiliation{Research and Education Center for Natural Sciences, Keio University, Hiyoshi 4-1-1, Yokohama, Kanagawa 223-8521, Japan}
\affiliation{Nishogakusha University, 6-16, Sanbancho, Chiyoda, Tokyo 102-8336, Japan}
\author{Muneto~Nitta}
\email[]{nitta@phys-h.keio.ac.jp}
\affiliation{Department of Physics and Research and Education Center for Natural Sciences, Keio University, 4-1-1 Hiyoshi, Kanagawa 223-8521, Japan}
\affiliation{International Institute for Sustainability with Knotted Chiral Meta Matter (SKCM$^2$), Hiroshima University, 1-3-2 Kagamiyama, Higashi-Hiroshima, Hiroshima 739-8511, Japan}
\author{Chihiro~Sasaki}
\email[]{chihiro.sasaki@uwr.edu.pl}
\affiliation{Institute of Theoretical Physics, University of Wroclaw, plac Maksa Borna 9, PL-50204 Wroclaw,
Poland}
\affiliation{International Institute for Sustainability with Knotted Chiral Meta Matter (SKCM$^2$), Hiroshima University, 1-3-2 Kagamiyama, Higashi-Hiroshima, Hiroshima 739-8511, Japan}


\date{\today}

\begin{abstract}
We propose novel superfluids induced by the parity-doubled baryons.
The parity-doubled baryons, i.e., a nucleon $N(940)$ with
mass 940 MeV and
 spin-parity $J^{P}=1/2^{+}$ and an excited nucleon $N^{\ast}(1535)$ with
 mass 1535 MeV and spin-parity
 $J^{P}=1/2^{-}$ in vacuum, become degenerate at sufficiently high density where the chiral symmetry is restored.
In this study, we extend the conventional $\U(1)$ chiral symmetry to the higher dimensional symmetries, 
dubbed emergent chiral symmetries,
including the naive and mirror assignments as their subgroups.
Starting with the Lagrangian up to four-point interactions among the neutron $n$ and its chiral partner $n^{\ast}$, neutral components in $N$ and $N^{\ast}$, in pure neutron matter, we investigate the properties of the ground state with a pairing gap generated by the $n$ and $n^{\ast}$ in the mean-field approximation.
We find vector-type condensates that induce the dynamical breaking of 
a new class of internal symmetries, {\it emergent chiral symmetries}, as well as 
the baryon number and the rotational symmetries of the real space, indicating the appearance of massless Nambu-Goldstone bosons consisting of six quarks: emergent pions, superfluid phonons, and magnons, respectively. 
We also study the fermionic excitation modes at low-energy scales, and show that they exhibit a spatial anisotropy of the propagation at the Dirac cone in momentum space.
Some phenomenological implications are advocated, shedding new light on the properties of neutron stars.
\end{abstract}

\keywords{Chiral symmetry, parity-doubled baryons, neutron matter, superfluidity, neutron stars}

\maketitle


\section{Introduction}
\label{sec:introduction}

Recent advances in multimessenger astronomy have provided a key set of constraints on the equation of state (EoS) of QCD at high baryon density,
via extraterrestrial observations of dense stellar objects such as neutron stars (NSs)~\cite{Annala:2023cwx}.
It was suggested~\cite{Fattoyev:2017jql} that the softening of the EoS at intermediate densities required to comply with the observed tidal deformability,
together with the subsequent stiffening at high densities to protect the $2 M_\odot$ NSs from their gravitational collapse,
may indicate a phase transition at the core.
This is often attributed to color deconfinement~\cite{Annala:2019puf,Christian:2021uhd,Li:2021sxb},
although it is rather a heuristic guidance within simplistic approaches that do not account for the microscopic dynamics of underlying QCD.

Whereas the existence of quark matter is an intriguing option, a pure hadronic scenario has been shown to fully reconcile the modern constraints
from observation~\cite{Marczenko:2018jui,Marczenko:2020jma,Minamikawa:2020jfj,Marczenko:2021uaj,Marczenko:2022hyt,Gao:2024chh}, where the essential ingredients are
the lowest nucleon $N(940)$
(mass 940 MeV and spin-parity $J^{P}=1/2^{+}$)
and its chiral partner $N^*(1535)$
(mass 1535 MeV and spin-parity $J^{P}=1/2^{-}$)
interacting with the scalar and pseudoscalar mesons in a chiral-invariant way.
The properties of $N$ and $N^*$ can be well captured within the parity-doublet model based on the {\it mirror assignment}
of chirality~\cite{Detar:1988kn,Jido:2001nt}:
the mirror assignment is a more general representation for the nucleons with opposite parity, than the conventional one \`a la Gell-Mann and Levy,
the naive assignment. It is a striking difference that the parity doublet model predicts the $N$ and $N^*$ being degenerate and {\it massive}
when the chiral symmetry becomes restored, whereas the naive one inevitably leads to massless nucleons.
The former, the mirror scenario, emerges in nature, as confirmed by the recent lattice QCD simulations exhibiting a clear manifestation of
parity doubling near the chiral crossover in the baryon spectra~\cite{Aarts:2015mma,Aarts:2017rrl,Aarts:2018glk}.
Thus, the hadronic matter at high density has recently attracted attention as an interesting realization of QCD substances in the confined phase
(see, e.g., a recent review Ref.~\cite{Minamikawa:2023eky}).

At high density, nucleons may form Cooper pairs giving rise to superfluidity.
Neutron superfluids have been studied intensively not only in astrophysics and nuclear physics but also in condensed matter physics~\cite{Chamel2017,Haskell:2017lkl,Sedrakian:2018ydt,Graber:2016imq,Baym:2017whm}.
It has been discussed since the early studies that the neutron $^{1}S_{0}$ (spin-singlet and $S$-wave) superfluids exist at low densities near the surface of neutron stars~\cite{Migdal:1960,Wolf1966}, and the neutron $^{3}P_{2}$ (spin-triplet and $P$-wave) superfluids exist at higher densities in the deep inside~\cite{Tabakin1968,Hoffberg1970,Tamagaki1970,Takatsuka1971,Takatsuka1972,Fujita1972,Richardson1972,
Sauls:1978lna,
Muzikar:1980as,
Sauls:1982ie,
Masuda:2015jka,
Mizushima:2016fbn,
Chatterjee:2016gpm,
Masuda:2016vak,
Mizushima:2017pma,
Yasui:2018tcr,
Masaki:2019rsz,
Mizushima:2019spl,
Yasui:2019tgc,
Yasui:2019vci,
Yasui:2019pgb,
Yasui:2019unp,
Yasui:2020xqb,
Masaki:2021hmk,
Mizushima:2021qrz,
Kobayashi:2022moc,
Kobayashi:2022dae,
Masaki:2023rtn,
Leinson:2020xjz,
Marmorini:2020zfp, 
Sedrakian:2024dgk}.
The neutron $^{3}P_{2}$ superfluids are tolerant against the strong magnetic fields such as in magnetars with $B=10^{15}$-$10^{18}$ G thus are important to understand the interior of neutron stars and magnetars.
In fact, it was studied that the enhancement of neutrino emission due to the neutron $^{3}P_{2}$ superfluids may explain the rapid cooling observed for the neutron star in Casiopeia A~\cite{Shternin2011,Page:2010aw,Heinke2010} (see also Refs.~\cite{Blaschke2012,Blaschke2013,Grigorian:2016leu}).
It was also discussed that the bosonic excitations in the neutron $^{3}P_{2}$ superfluids may be relevant for the cooling process by neutrino emission~\cite{Bedaque:2003wj,Leinson:2011wf,Leinson:2012pn,Leinson:2013si,Bedaque:2012bs,bedaquePRC14,Bedaque:2013fja,Bedaque:2014zta,Leinson:2009nu,Leinson:2010yf,Leinson:2010pk,Leinson:2010ru,Leinson:2011jr}.
It has been proposed that vortices penetrating  $^3P_2$-$^1S_0$ interfaces inside the neutron stars can explain pulsar glitches 
without fine-tuning~\cite{Marmorini:2020zfp}
(see also, e.g., a recent review Ref.~\cite{Antonelli:2023vpd}).\footnote{It has been discussed that, beyond glitches, superfluidity might also influence pulsar timing noise in neutron stars~\cite{Antonelli:2022gqw}.} 
In the condensed-matter aspects,
it is particularly interesting that the $^{3}P_{2}$ superfluids are topological superfluids, 
admitting topologically protected Majorana fermions on 
their surfaces~\cite{Mizushima:2016fbn} 
and inside cores~\cite{Masaki:2019rsz,Masaki:2021hmk}
of singly quantized vortices 
\cite{Muzikar:1980as,Sauls:1982ie,
Masuda:2015jka} and
half-quantum vortices~\cite{Masuda:2016vak,
Kobayashi:2022moc,
Kobayashi:2022dae} 
(see Ref.~\cite{Masaki:2023rtn} as a review),
and thus they may be called the largest topological matter in the Universe.

In the past studies, it was conventionally assumed that the chiral symmetry is spontaneously broken in the neutron $^{1}S_{0}$ and $^{3}P_{2}$ superfluids as in vacuum
(see, e.g., a review Ref.~\cite{Sedrakian:2018ydt}).
However, it is expected that the chiral symmetry can become restored at sufficiently high density inside neutron stars even if the substances remain composed of hadrons, such as in the parity-doublet model.
For an example, it has been discussed that the baryon number density of hadronic matter could reach maximally $3$-$5n_{0}$ ($n_{0}=0.16$ fm$^{-3}$: normal nuclear matter density) (see, e.g.,  Refs.~\cite{Marczenko:2018jui,Marczenko:2020jma,Minamikawa:2020jfj} and references therein).
It is of particular interest to explore the properties of neutron superfluids embracing potential modifications of hadrons due to (partially) restored chiral symmetry.

\begin{figure}
\includegraphics[keepaspectratio, scale=0.2]{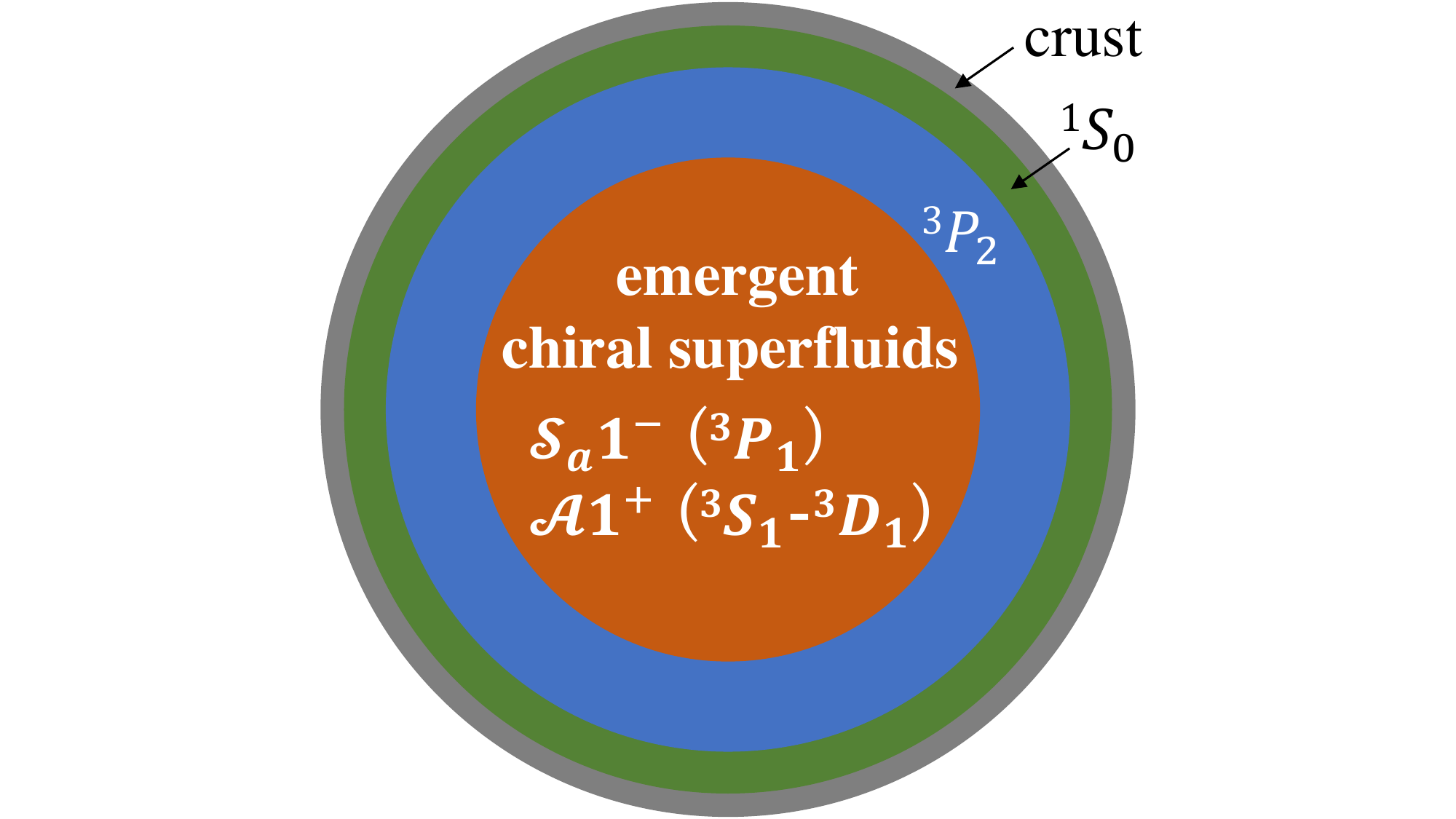} %
\caption{Schematic cross section inside a neutron star formed by the crust, the neutron $^{1}S_{0}$ superfluids near the surface, the neutron $^{3}P_{2}$ superfluids at intermediate region, and the
emergent chiral superfluids
in the deep interior. Phases of the
emergent chiral superfluids
are denoted by ${\cal S}_{a}1^{-}$ ($^{3}P_{1}$) ($a=0,1,3$) and ${\cal A}1^{+}$ ($^{3}S_{1}$-$^{3}D_{1}$).
Here, ${\cal S}_{a}$ and ${\cal A}$ indicate the symmetric and antisymmetric representations in the emergent chiral symmetry, and $J^{P}$ represents total angular momentum $J$ and parity $P$.
 In the parentheses, the nonrelativistic quantum numbers, corresponding to $J^{P}=1^{-}$ and $1^{+}$, are denoted by $^{2S+1}L_{J}$ with spin $S$ and angular momentum $L$.}
\label{fig:240807_star_inside}
\end{figure}

In this article, we propose superfluidity of parity-doubled neutrons for the first time.
This new type of superfluid is characterized by the Cooper pairings given not only by $nn$ but also $nn^{\ast}$ and $n^{\ast}n^{\ast}$,
where $n^{\ast}$ (spin-parity $J^{P}=1/2^{-}$) is an excited state of a neutron $n$, i.e., the neutral state of $N^{\ast}$.
It is natural to introduce an
internal symmetry for the two fermions in a doublet, 
in particular an exchange symmetry 
between them, since there is no distinction between them {\it a priori};
in addition to the conventional chiral symmetry, 
the doublet can have an extra symmetry, either  
$\U(1)_{(1-2)\L} \times \U(1)_{(1-2)\R}$ 
(the relative phase of the doublet) 
or
$\SU(2)_{\L} \times \SU(2)_{\R}$,
dubbed an {\it emergent chiral symmetry}.
The emergent chiral symmetries include the naive and mirror assignments as subgroups, 
thus providing us  in-depth understanding of the chiral symmetry.
We will construct the full Lagrangian including four-point interactions among the neutron $n$ and its chiral partner $n^{\ast}$ and present the phase structure in the mean field approximation.
We find novel vector-type $J^{P}=1^{+}$ and $1^{-}$ condensates 
($^{3}P_{1}$ and $^{3}S_{1}$-$^{3}D_{2}$ in nonrelativistic notation)
exhibiting 
the dynamical breaking of 
the emergent chiral symmetries, 
the baryon number symmetry, 
and the rotational symmetry of the real space, 
resulting in 
emergent pions, superfluid phonons, and magnons, respectively as gapless Nambu-Goldstone (NG) bosons consisting of six quarks.
We also find low-energy fermionic excitation modes exhibiting a spatial anisotropy of the propagation.
We shall name those new states of superfluidity {\it emergent chiral superfluids}.
The emergent chiral superfluids may exist in a more inner part of NSs than the phases of neutron $^{1}S_{0}$ and $^{3}P_{2}$ superfluids 
as illustrated in Fig.~\ref{fig:240807_star_inside}.
It has been shown numerically that the neutron $^{3}P_{2}$ pairings can exist at baryon number densities up to $3n_{0}$ or a few more (see, e.g., Refs.~\cite{Tamagaki1970,Takatsuka1971,Takatsuka1972}).
Therefore, it should be expected that the baryon number densities relevant to the emergent chiral superfluids would be larger than those densities.\footnote{In the present study, we consider the case that the critical density for the restoration of chiral symmetry should be higher than the density of $^{3}P_{2}$ superfluids.}

The paper is organized as follows:
in Sec.~\ref{sec:formalism}, we introduce the emergent chiral symmetry for the parity-doubled neutrons as an extension of the usual chiral symmetry.
Based on the emergent chiral symmetry, we construct the Lagrangian with four-point interactions for the chiral-doublet baryons.
In Sec.~\ref{sec:phase_diagram}, we adopt the Namb-Gor'kov formalism to the particle-particle (hole-hole) interaction in the parity-doubled baryons and apply the mean field approximation by introducing gap functions. We show that the vector-type condensate accompanies the breaking of the emergent chiral symmetry, baryon number symmetry, and rotational symmetry in the ground state.
We also analyze the fermionic excitations in low-energy scale and show the existence of the Dirac points in the momentum space.
The final section is devoted to our conclusion and outlook.
In Appendix~\ref{sec:extension_chiral_symmetry}, the naive and mirror assignments are briefly reviewed along with the emergent chiral symmetry.
In Appendix~\ref{sec:Lagrangians_naive_mirror}, the Lagrangians in the two assignments are given for the simplest emergent chiral symmetry.
In Appendix~\ref{sec:Fierz_transformation}, the Fierz identities are summarized for convenience of readers.
In Appendix~\ref{sec:H3pm_DP_eigenfunctions}, the eigenfunctions of the Hamiltonian for the massless Dirac fermions are summarized.

\section{Symmetries and the Lagrangian} \label{sec:formalism}

\subsection{Naive and mirror assignments} \label{sec:naive_mirror}

To describe consistently a system for baryons both in the broken and restored phases of chiral symmetry,
we introduce parity-doubled baryons which are composed of a pair of a nucleon $N$ with spin-parity $J^{P}=1/2^{+}$ and an excited nucleon $N^{\ast}(1535)$ with $J^{P}=1/2^{-}$~\cite{DeTar_1989}.
The parity-doublet model has been applied to the hadron spectroscopy related to the chiral dynamics, see, e.g., Ref.~\cite{Jido:2001nt} for a review.
In this framework, the nucleon $N$ and the excited nucleon $N^{\ast}(1535)$ are expressed as a linear combination of two baryons denoted by $\Psi_{1}$ and $\Psi_{2}$ in the flavor eigenstates.
Assigning the chiralities to $\Psi_{1}$ and $\Psi_{2}$ ($N$ and $N^{\ast}$), two different types of chiral symmetry have been considered: a naive assignment and a mirror assignment~\cite{DeTar_1989}.
In the naive assignment, the left (right) component of $\Psi_{1}$ and the left (right) component of $\Psi_{2}$ rotate simultaneously.
In the mirror assignment, in contrast, the left (right) component of $\Psi_{1}$ and the right (left) component of $\Psi_{2}$ rotate simultaneously.
Notice the different combinations of the left and right components for $\Psi_{1}$ and $\Psi_{2}$, respectively, in the mirror assignment.~\footnote{A parity doublet has also been referred to as a chiral doublet in the literature, e.g.~\cite{Nowak:1992um,Bardeen:1993ae,Nowak:2003ra,Harada:2003kt}.}

To model a neutron superfluid in the context of parity doublet framework, we shall consider the interaction between two parity-doubled baryons, i.e., $N$ and $N^{\ast}$.
Because our main interest is primarily put on the neutron stars, we consider only the neutral components $n$ and $n^{\ast}$ by neglecting the charged ones.
Here $n^{\ast}$ indicates $n^{\ast}(1535)$, i.e., the neutral component of $N^{\ast}(1535)$, for shorthand notation.
Thus, $n$ and $n^{\ast}$ are expressed as a linear combination of $\psi_{1}$ and $\psi_{2}$:
$n = \cos \theta \psi_{1} + \gamma_{5} \sin \theta \psi_{2}$ and $n^{\ast} = -\gamma_{5} \sin \theta \psi_{1} + \cos \theta \psi_{2}$ with an appropriate mixing angle $\theta$~\cite{Detar:1988kn,Jido:2001nt}.
Here we consider that $\psi_{1}$ and $\psi_{2}$ have only neutral components in $\Psi_{1}$ and $\Psi_{2}$, respectively.
Characterizing our superfluid, we have $nn$, $nn^{\ast}$ and $n^{\ast}n^{\ast}$ interactions to get the particle-particle (hole-hole) pairings.
We will consider a full set of four-point interactions (zero-range force) for $n$ and $n^{\ast}$.
This setup is much different from the previous treatments of the parity-doublet model~\cite{Marczenko:2018jui,Marczenko:2020jma,Minamikawa:2020jfj,Marczenko:2021uaj,Marczenko:2022hyt}, where the $\sigma$ and $\omega$ fields are introduced to couple to the parity-doubled baryons like in the relativistic mean-field model of nuclear matter~\cite{Serot:1984ey}.

In order to constrain the interaction terms for $n$ and $n^{\ast}$
(or $\psi_{1}$ and $\psi_{2}$),
we shall elevate the usual chiral symmetry to the {\it emergent chiral symmetry} as explained in the next subsection.
This is the novel symmetry which is proposed in this study for the first time.
The emergent chiral symmetry can be regarded as a proper symmetry, including the naive and mirror assignments as subgroups, in the Hamiltonian of the parity doublet model.
In this study, we consider two different options: $\U(1)_{(1-2)\L} \times \U(1)_{(1-2)\R}$
and $\SU(2)_{\L} \times \SU(2)_{\R}$ emergent chiral symmetries. 
To the best of our knowledge, such extension of the parity doublet has never been considered in the literature, deserving to explore the properties of the parity-doublet model in a broader framework.
This study is especially devoted to the first application of the new parity-doublet model to the superfluidity in neutron matter.

\subsection{Emergent chiral symmetry} 

\label{sec:emergent chiral_symmetry}

Given our motivation that the two options of chirality assignments are to be unified,
we extend the usual chiral symmetry to higher symmetry: an emergent chiral symmetry.
Among possible candidates, we are naturally led to 
the
$\U(1)_{1\L} \times \U(1)_{2\L} \times \U(1)_{1\R} \times \U(1)_{2\R}$
symmetry, since there are four fermions 
$\psi_{i\alpha}$ ($i=1, 2$ and $\alpha=\L, \R$)
in the parity-doubled baryons including its chirality, 
on which four phase degrees of freedom $\U(1)_{i\alpha}$ act.
First, let us define 
the ordinary chiral symmetry 
$\U(1)_{\L} \times \U(1)_{\R}$ 
as the overall phase transformation
and the (simplest) emergent chiral symmetry
$G_{\rm L} \times G_{\rm R} \equiv \U(1)_{(1-2)\L} \times  \U(1)_{(1-2)\R}$ as
an opposite phase rotation of the two fermions $\psi_{1\alpha}$ and $\psi_{2\alpha}$:
\begin{align}
 \psi_{\L} \to e^{i\theta_{\L}} \psi_{\L}, 
  \quad 
  \psi_{\R} \to e^{i\theta_{\R}} \psi_{\R},
   \quad &{\rm with} \quad
  (e^{i\theta_{\L}},
  e^{i\theta_{\R}})\in 
   \U(1)_{\L} \times \U(1)_{\R},
   \\ 
 \psi_{\L} \rightarrow e^{i\tau_{3}\theta_{\L}}\psi_{\L}, \quad
\psi_{\R} \rightarrow e^{i\tau_{3}\theta_{\R}}\psi_{\R}, \quad & {\rm with} \quad
  (e^{i\tau_{3}\theta_{\L}} , 
  e^{i\tau_{3}\theta_{\R}})
  \in  
  \U(1)_{(1-2)\L} \times  \U(1)_{(1-2)\R} , 
\end{align}
respectively.~\footnote{Here $\tau_{0}$, $\tau_{1}$, $\tau_{2}$ and $\tau_{3}$ are the Paul matrices given by
\begin{align}
\tau_{0}=
\begin{pmatrix}
1 & 0 \\
0 & 1
\end{pmatrix}, \,
\tau_{1}=
\begin{pmatrix}
0 & 1 \\
1 & 0
\end{pmatrix}, \,
\tau_{2}=
\begin{pmatrix}
0 & -i \\
i & 0
\end{pmatrix}, \,
\tau_{3}=
\begin{pmatrix}
1 & 0 \\
0 & -1
\end{pmatrix}.
\end{align}
}
Here we have defined $\psi_{\L}^{\t}=(\psi_{1\L},\psi_{2\L})^{\t}$ and $\psi_{\R}^{\t}=(\psi_{1\R},\psi_{2\R})^{\t}$,
where $\psi_{\alpha}$ is the doublet for each chirality ($\alpha=\L,\R$).
Then, 
we notice that the $\U(1)_{1\L} \times \U(1)_{2\L}$ and $\U(1)_{1\R} \times \U(1)_{2\R}$ symmetries 
can be expressed in terms of these symmetries as
\begin{align}
   \U(1)_{1\L} \times \U(1)_{2\L}
=
   \frac{\U(1)_{\L} \times \U(1)_{(1-2)\L}}{{\mathbb Z}_{2\L}^\prime},
\quad
   \U(1)_{1\R} \times \U(1)_{2\R}
=
   \frac{\U(1)_{\R} \times \U(1)_{(1-2)\R}}{{\mathbb Z}_{2\R}^\prime},
\label{eq:U1_U1_symmetry_decomposition}
\end{align}
for each chirality.  
The discrete group ${\mathbb Z}_{2\alpha}^\prime$ 
in the denominators 
on the right-hand sides 
is necessary, because a simultaneous action 
of the $\pi$ rotation  
$\psi_{\alpha} \rightarrow e^{i\tau_{3}\pi}\psi_{\alpha} = - \psi_{\alpha}$ 
of $\U(1)_{(1-2)\alpha}$
and that 
$e^{i\pi}\psi_{\alpha} = - \psi_{\alpha}$ 
of 
$\U(1)_{\alpha}$ 
is redundant and should be removed. 
In this paper, we shall name
the $\U(1)_{(1-2)\L} \times \U(1)_{(1-2)\R}$ 
symmetry 
{\it an emergent chiral symmetry}.

The ordinary chiral symmetry $\U(1)_{\L} \times \U(1)_{\R}$ is of course
a subgroup of the 
total global symmetry by definition
\begin{align}
   \U(1)_{\L} \times \U(1)_{\R}
\subset
   \frac{\U(1)_{\L} \times \U(1)_{(1-2)\L}}{{\mathbb Z}_{2\L}^\prime}
   \times
   \frac{\U(1)_{\R} \times \U(1)_{(1-2)\R}}{{\mathbb Z}_{2\R}^\prime}
 = \U(1)_{1\L} \times \U(1)_{2\L} \times \U(1)_{1\R} \times \U(1)_{2\R}.
   \label{eq:chiral0}
\end{align}
The $\U(1)_{(1-2)\L} \times \U(1)_{(1-2)\R}$ emergent chiral symmetry
expresses the internal symmetry for the doublet $(\psi_{1},\psi_{2})^{\t}$.
This is a novel symmetry which has not been recognized in the past studies~\cite{Detar:1988kn,Jido:2001nt}.
The $\U(1)_{(1-2)\L} \times \U(1)_{(1-2)\R}$ emergent chiral symmetry reproduces the naive assignment and the mirror assignment as the subgroups
when a certain symmetry-locking is introduced: if our theory respects only 
a simultaneous phase rotation of 
$\U(1)_{1\L}$ and $\U(1)_{2\L}$ and that of $\U(1)_{1\R}$ and $\U(1)_{2\R}$
it gives the naive assignment, 
while if it does only 
that of $\U(1)_{1\L}$ and $\U(1)_{2\R}$  
and that of $\U(1)_{2\L}$ and $\U(1)_{1\R}$, 
it gives the mirror assignment.
Therefore, the naive and mirror assignments can be identified as the subgroups $\U(1)_{1\L+2\L} \times \U(1)_{1\R+2\R}$ and $\U(1)_{1\L+2\R} \times \U(1)_{1\R+2\L}$, respectively.
In this sense, the $\U(1)_{(1-2)\L} \times \U(1)_{(1-2)\R}$ emergent chiral symmetry will provide us with a unified picture for the parity-doubled baryons.
Given that the mirror assignment is more favored than the naive assignment in the recent lattice QCD simulations at finite temperature~\cite{Aarts:2015mma,Aarts:2017rrl,Aarts:2018glk}~\footnote{As a related work, the parity-doublet model is adopted also for the fluctuations around the critical point in the QCD phase diagram~\cite{Koch:2023oez}.},
we may naturally ask how the emergent chiral symmetry, as an extension of the usual chiral symmetry, can appear in a variety of QCD phases at finite temperature, finite density, external fields, and so on.
To this end, we shall study how the emergent chiral symmetry arises in various phenomena of hadronic matter.

\begin{figure}
\includegraphics[keepaspectratio, scale=0.25]{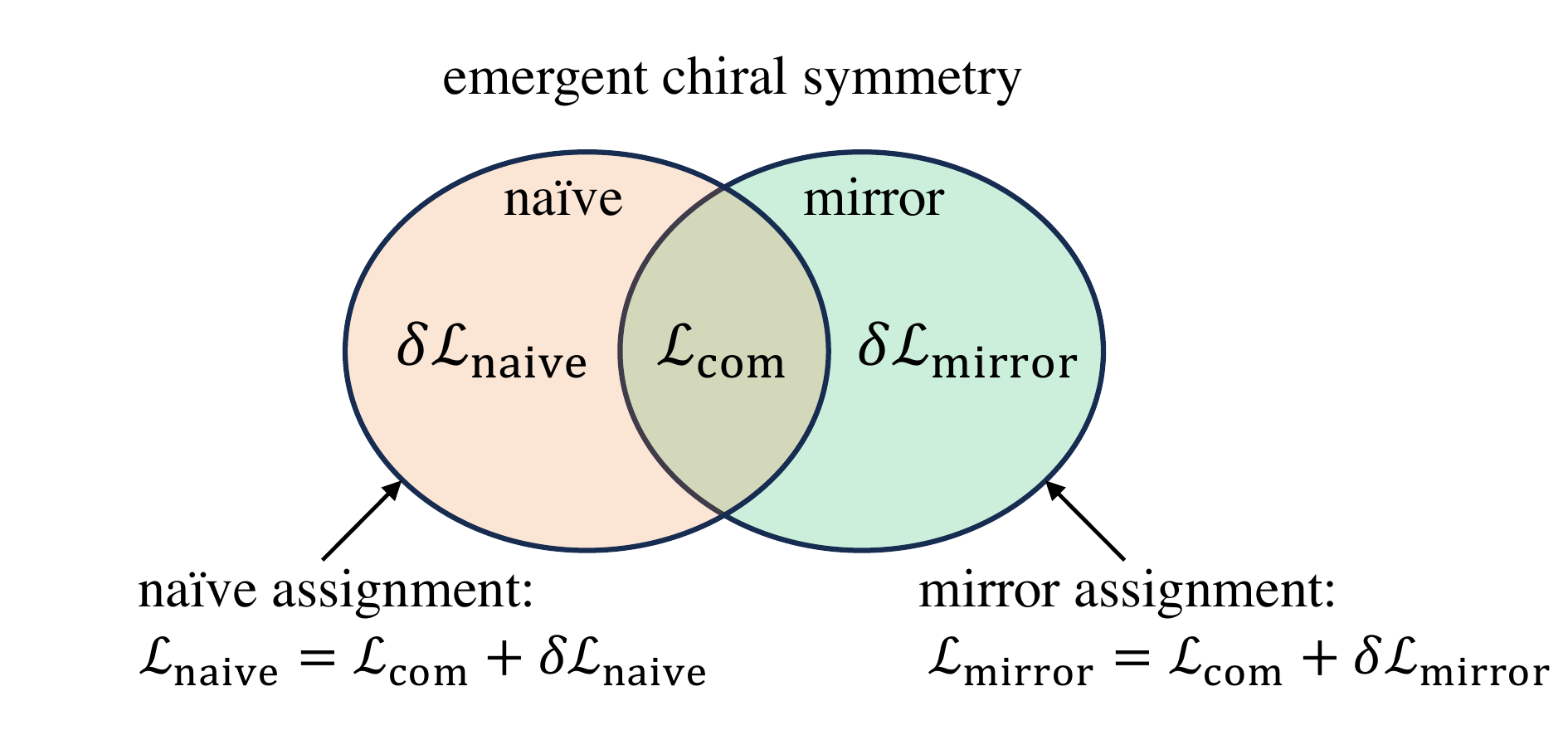} %
\caption{
The schematic figure for the inclusion relations among the Lagrangian terms, ${\cal L}_{\mathrm{com}}$, ${\cal L}_{\mathrm{naive}}$ and ${\cal L}_{\mathrm{mirror}}$, is shown.}
\label{fig:240827_common_figure}
\end{figure}

With the $\U(1)_{(1-2)\L} \times \U(1)_{(1-2)\R}$ emergent chiral symmetry, the most general Lagrangian
is expressed as a sum of three terms as
\begin{align}
   {\cal L} = {\cal L}_{\mathrm{com}} + \delta {\cal L}_{\mathrm{naive}} + \delta {\cal L}_{\mathrm{mirror}}.
\end{align}
Here ${\cal L}_{\mathrm{com}}$ is the one including only the terms invariant commonly in the naive and mirror assignments.
In contrast, $\delta {\cal L}_{\mathrm{naive}}$ ($\delta {\cal L}_{\mathrm{mirror}}$) is the term invariant for the naive (mirror) assignment as a subgroup of the $\U(1)_{(1-2)\L} \times \U(1)_{(1-2)\R}$ emergent chiral symmetry, but it is composed of only the terms which are not included in ${\cal L}_{\mathrm{com}}$.
The Lagrangians invariant in the naive and mirror assignments are given by ${\cal L}_{\mathrm{naive}} = {\cal L}_{\mathrm{com}} + \delta {\cal L}_{\mathrm{naive}}$ and ${\cal L}_{\mathrm{mirror}} = {\cal L}_{\mathrm{com}} + \delta {\cal L}_{\mathrm{mirror}}$, respectively. See also Fig.~\ref{fig:240827_common_figure}.
Note that $\delta {\cal L}_{\mathrm{naive}}$ and $\delta {\cal L}_{\mathrm{mirror}}$ break the symmetry of ${\cal L}_{\mathrm{com}}$ explicitly.
In the present study, we will focus on the ${\cal L}_{\mathrm{com}}$ to clarify generic properties shared by the naive and mirror assignments as a first step of our study.

To describe the emergent chiral superfluids, we introduce the four-point interaction for $\psi_{1}$ and $\psi_{2}$.
Thus we consider the Lagrangian invariant
under the $\U(1)_{(1-2)\L} \times \U(1)_{(1-2)\R}$ emergent chiral symmetry
\begin{align}
   {\cal L}_{\mathrm{com}}^{\U(1)_{(1-2)\L} \times \U(1)_{(1-2)\R}}
&=
   \bar{\psi}_{1} i \gamma\partial \psi_{1} + \bar{\psi}_{2} i \gamma\partial \psi_{2}
   \nonumber \\ & 
 - 4g_{\perp}
   \bigl( (\bar{\psi}_{1}\psi_{1})^{2} + (\bar{\psi}_{1}i\gamma_{5}\psi_{1})^{2} \bigr)
 - 4g_{\perp}'
   \bigl( (\bar{\psi}_{2}\psi_{2})^{2} + (\bar{\psi}_{2}i\gamma_{5}\psi_{2})^{2} \bigr)
   \nonumber \\ & 
 - 8g_{\parallel}
   \bigl( (\bar{\psi}_{1}\psi_{2})(\bar{\psi}_{2}\psi_{1}) + (\bar{\psi}_{1}i\gamma_{5}\psi_{2})(\bar{\psi}_{2}i\gamma_{5}\psi_{1}) \bigr).
\label{eq:Lagrangian_general_summary}
\end{align}
The invariance of ${\cal L}_{\mathrm{com}}$ holds also for both of the naive and mirror assignments.
Here we have three interaction terms with the coupling constants, $g_{\perp}$, $g_{\perp}'$, and $g_{\parallel}$, which take different values in general.
The coefficients are multiplied by constant numbers, $-4$ and $-8$, for the later convenience.
Similarly, we can construct the Lagrangians ${\cal L}_{\mathrm{naive}}$ and ${\cal L}_{\mathrm{mirror}}$ for the naive and mirror assignments, respectively, as shown in Eqs.~\eqref{eq:Lagrangian_naive_summary} and \eqref{eq:Lagrangian_mirror_summary} in Appendix~\ref{sec:Lagrangians_naive_mirror}.
In this study, we rather concentrate on the ${\cal L}_{\mathrm{com}}$ to clarify the common properties shared by the naive and mirror assignments.

Besides the $\U(1)_{\L} \times \U(1)_{\R}$ chiral symmetry,
it is natural to
introduce a further internal symmetry for the two fermions $\psi_{1}$ and $\psi_{2}$ to impose additional constraints in the Lagrangian~\eqref{eq:Lagrangian_general_summary}.
The requirement of such an internal symmetry stems from the observation that there is no {\it a priori} criterion to distinguish $\psi_{1}$ and $\psi_{2}$ in essence, and hence some internal symmetry may naturally arise to exchange $\psi_{1}$ and $\psi_{2}$.
Referring to the decompositions in Eq.~\eqref{eq:U1_U1_symmetry_decomposition},
the above consideration leads to  enlargement of 
the 
emergent chiral symmetry,
$\U(1)_{(1-2)\L} \times \U(1)_{(1-2)\R}$,
to higher symmetry to realize the exchange of $\psi_{1}$ and $\psi_{2}$.
The higher emergent chiral symmetry represents the additional constraint for the exchange of $\psi_{1}$ and $\psi_{2}$ reduces the number of
terms in the Lagrangian~\eqref{eq:Lagrangian_general_summary}.

As an exchange symmetry 
between $\psi_{1}$ and $\psi_{2}$ in the doublet $\psi_{\alpha}^{\t}=(\psi_{1\alpha},\psi_{2\alpha})^{\t}$ for each chirality, 
one can consider 
\begin{align}
G_{\L} \times G_{\R} = \SU(2)_{\L} \times \SU(2)_{\R},
\label{eq:emergent-chiral2}
\end{align}
with an $\SU(2)$ symmetry continuously exchanging $\psi_{1}$ and $\psi_{2}$ for each chirality:
$\psi_{\alpha} \rightarrow e^{i\vec{\tau}\cdot\vect{\theta}_{\alpha}/2} \psi_{\alpha}$ ($\alpha=\L, \R$) 
with 
the Pauli matrices, $\vec{\tau}=(\tau_{1},\tau_{2},\tau_{3})$, and the real parameters, $\vect{\theta}_{\alpha}=(\theta_{1\alpha},\theta_{2\alpha},\theta_{3\alpha})$.

\begin{figure}
\includegraphics[keepaspectratio, scale=0.25]{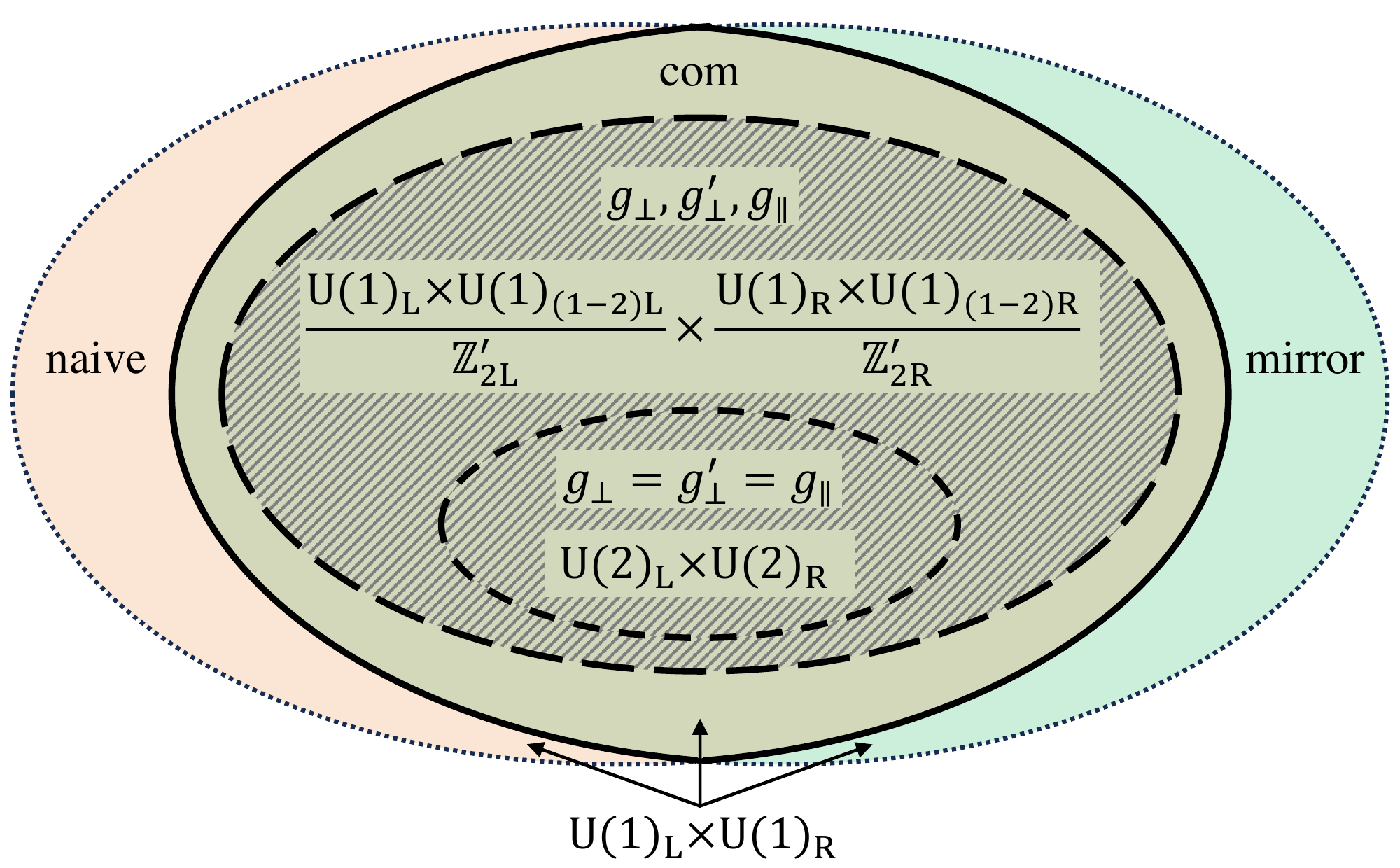} %
\caption{
Inclusion relations are shown for the Lagrangians with 
$\frac{\U(1)_{\L} \times \U(1)_{(1-2)\L}}{{\mathbb Z}_{2\L}^\prime}
\times
\frac{\U(1)_{\R} \times \U(1)_{(1-2)\R}}{{\mathbb Z}_{2\R}^\prime}$
for the $\U(1)_{(1-2)\L} \times \U(1)_{(1-2)\R}$ emergent chiral symmetry 
and $\U(2)_{\L} \times \U(2)_{\R}$ for the $\SU(2)_{\L} \times \SU(2)_{\R}$ emergent chiral symmetry.
In the present study, we consider the Lagrangian terms commonly shared by the naive and mirror assignments as indicated by solid lines (com).
See also Fig.~\ref{fig:240827_common_figure}.
$\U(1)_{\L} \times \U(1)_{\R}$ indicates
$\U(1)_{1\L+2\L} \times \U(1)_{1\R+2\R}$ in the naive assignment and 
$\U(1)_{1\L+2\R} \times \U(1)_{1\R+2\L}$ in the mirror assignment.
The $\U(1)_{\L} \times \U(1)_{1\R}$ symmetric term included commonly in both naive and mirror assignments but possessing no $\frac{\U(1)_{\L} \times \U(1)_{(1-2)\L}}{{\mathbb Z}_{2\L}^\prime}
\times
\frac{\U(1)_{\R} \times \U(1)_{(1-2)\R}}{{\mathbb Z}_{2\R}^\prime}$ symmetry is shown by the $g_{1212}$ term in Eqs.~\eqref{eq:Lagrangian_naive_summary} and \eqref{eq:Lagrangian_mirror_summary} in Appendix~\ref{sec:Lagrangians_naive_mirror}.}
\label{fig:240902_inclusion}
\end{figure}

Since the emergent chiral symmetries 
$G_{\L} \times G_{\R}$ 
are extensions of 
$ \U(1)_{(1-2)\L}
   \times
 \U(1)_{(1-2)\R}$, their actions have 
a redundancy with that of 
the chiral symmetry 
 $\U(1)_{1\R} \times \U(1)_{2\R}$ that should be removed as 
 in Eq.~\eqref{eq:chiral0}
\begin{align}
    \frac{\U(1)_{\L} \times G_{\L}  }{{\mathbb Z_{2\L}^\prime}}
   \times \frac{\U(1)_{\R} \times G_{\R}}{{\mathbb Z}_{2\R}^{\prime}} 
= \left\{
\renewcommand{\arraystretch}{2}
\begin{matrix}
  \displaystyle{
  \frac{\U(1)_{\L} \times \U(1)_{(1-2)\L}}{{\mathbb Z}_{2\L}^\prime}
  \times \frac{\U(1)_{\R} \times \U(1)_{(1-2)\R}}{{\mathbb Z}_{2\R}^{\prime}} 
  }
   &
   \\  
   \displaystyle{
   \frac{\U(1)_{\L} \times \SU(2)_{\L}}{{\mathbb Z}_{2\L}^{\prime}}
   \times \frac{\U(1)_{\R} \times \SU(2)_{\R}}{{\mathbb Z}_{2\R}^{\prime}}
   }
& \displaystyle{
   = \U(2)_{\L} \times \U(2)_{\R}
   }.
\end{matrix}
\renewcommand{\arraystretch}{1}
\right.
\label{eq:emergent-chiral3}
\end{align}
Here ${\mathbb Z}_{2\alpha}^{\prime}$ 
is the same one as defined in 
Eq.~\eqref{eq:U1_U1_symmetry_decomposition}.
These symmetries satisfy the following 
sequential inclusion relations:
\begin{align}
 &
   \U(1)_{\L} \times \U(1)_{\R}
\nonumber \\ 
&
\subset
   \frac{\U(1)_{\L} \times \U(1)_{(1-2)\L} }{
   {\mathbb Z}_{2\L}^\prime}
   \times 
   \frac{\U(1)_{\R} \times \U(1)_{(1-2)\R}}{
   {\mathbb Z}_{2\R}^\prime}
=
 \U(1)_{1\L} \times \U(1)_{2\L} \times \U(1)_{1\R} \times \U(1)_{2\R}
\nonumber \\ 
&
\subset
    \frac{\U(1)_{\L} \times \SU(2)_{\L}}{{\mathbb Z_{2\L}^\prime}}
   \times \frac{\U(1)_{\R} \times \SU(2)_{\R}}{{\mathbb Z}_{2\R}^\prime}
=
   \U(2)_{\L} \times \U(2)_{\R},
\label{eq:symmetry_inclusion}
\end{align}
where  the original chiral symmetry 
$\U(1)_{\L} \times \U(1)_{\R}$ 
in 
the first line 
 is extended into
the $G_{\L} \times G_{\R} = \U(1)_{(1-2)\L} \times \U(1)_{(1-2)\R} $
and $ \SU(2)_{\L} \times \SU(2)_{\R}$ 
emergent chiral symmetries
 (see Table~\ref{tbl:Z2_SU2PF} and Fig.~\ref{fig:240902_inclusion}).
In the following, we study the consequences of 
the 
$\U(1)_{(1-2)\L} \times \U(1)_{(1-2)\R}$
and $\SU(2)_{\L} \times \SU(2)_{\R}$ 
emergent chiral symmetries and their thermodynamics.

As relevant studies, we comment that the
$\U(N) \times \U(N)$  
chiral symmetric 
Gross-Neveu models 
were carried out in the $1+1$ dimensional systems in Refs.~\cite{Takahashi:2015nda,Thies:2015dim,Thies:2023dyh}.

\begin{table}[tp]
\begin{center}
\renewcommand{\arraystretch}{1.5}
\begin{tabular}{ll}
 \hline
 Full chiral symmetry & Transformation rule ($\alpha=\L, \R$) \\ \hline
 $\displaystyle
 \frac{\U(1)_{\L} \times \U(1)_{(1-2)\L}}{{\mathbb Z}_{2\L}^\prime}
 \times \frac{\U(1)_{\R} \times \U(1)_{(1-2)\R}}{{\mathbb Z}_{2\R}^\prime}
 $
 &
 $\psi_{\alpha} \rightarrow e^{i\varphi_{\alpha}} e^{i\tau_{3}\theta_{3\alpha}/2}\psi_{\alpha}$
 \\
 $\U(2)_{\L} \times \U(2)_{\R}$ &
 $\psi_{\alpha} \rightarrow e^{i\varphi_{\alpha}} e^{i\vec{\tau}\cdot\vect{\theta}_{\alpha}/2}\psi_{\alpha}$ \\
 \hline
\end{tabular}
\renewcommand{\arraystretch}{1}
\end{center}
\caption{Full chiral symmetries are summarized:
 $\frac{\U(1)_{\L} \times \U(1)_{(1-2)\L}}{{\mathbb Z}_{2\L}^\prime}
 \times \frac{\U(1)_{\R} \times \U(1)_{(1-2)\R}}{{\mathbb Z}_{2\R}^\prime}$
 and
 $\U(2)_{\L} \times \U(2)_{\R}$
 for the $\U(1)_{(1-2)\L} \times \U(1)_{(1-2)\R}$ and $\SU(2)_{\L} \times \SU(2)_{\R}$ emergent chiral symmetries, respectively, 
where 
 the actions of their universal covering groups
 $\U(1)_{\L} \times \U(1)_{(1-2)\L} \times \U(1)_{\R} \times \U(1)_{(1-2)\R}$
 and
 $\U(1)_{\L} \times 
 \SU(2)_{\L} \times 
 \U(1)_{\R} \times 
 \SU(2)_{\R} $
 are shown.
 $\psi_{\alpha}^{\t}=(\psi_{1\alpha},\psi_{2\alpha})^{\t}$ is the doublet ($\alpha=\L, \R$).
 $\vec{\tau}=(\tau_{1},\tau_{2},\tau_{3})$ are the Pauli matrices in the $\SU(2)$ space,
 $\varphi_{\alpha}$ is a real number, and
 $\vect{\theta}_{\alpha}=(\theta_{1\alpha},\theta_{2\alpha},\theta_{3\alpha})$
 are real parameters.
}
\label{tbl:Z2_SU2PF}
\end{table}%

\subsection{Lagrangians with the emergent chiral symmetries} \label{sec:Lagrangains_OS_SU2}

In this work, we will study the properties of the Lagrangian~\eqref{eq:Lagrangian_general_summary} by considering
the 
$\U(1)_{(1-2)\L} \times \U(1)_{(1-2)\R}$
and
$\SU(2)_{\L} \times \SU(2)_{\R}$ emergent chiral symmetries.
To start with under the $\U(1)_{(1-2)\L} \times \U(1)_{(1-2)\R}$ 
emergent chiral symmetry, we take a simplified Lagrangian
\begin{align}
   {\cal L}_{\mathrm{com}}^{\U(1)_{(1-2)\L} \times \U(1)_{(1-2)\R}\times {\mathbb Z}_2}
&=
   \bar{\psi} i \gamma\partial \psi
 - 2g_{\perp}
   \Bigl(
         \bigl( \bar{\psi}\tau_{0}\psi \bigr)^{2}
      + \bigl( \bar{\psi}\tau_{3}\psi \bigr)^{2}
      + \bigl( \bar{\psi}i\gamma_{5}\tau_{0}\psi \bigr)^{2}
      + \bigl( \bar{\psi}i\gamma_{5}\tau_{3}\psi \bigr)^{2}
   \Bigr)
   \nonumber \\ & 
 - 2g_{\parallel}
   \Bigl(
         \bigl(\bar{\psi}\tau_{1}\psi\bigr)^{2}
      + \bigl(\bar{\psi}\tau_{2}\psi\bigr)^{2}
      + \bigl( \bar{\psi}i\gamma_{5}\tau_{1}\psi \bigr)^{2}
      + \bigl( \bar{\psi}i\gamma_{5}\tau_{2}\psi \bigr)^{2}
   \Bigr),
\label{eq:Lagrangian_general_Z2}
\end{align}
where $g_{\perp}$ and $g_{\parallel}$ are the relevant coupling constants ($g_{\perp}=g'_{\perp}$) and
$\tau_{0}$ is a $2\times2$ unit matrix,
with the assumption that $g_{\perp}=g'_{\perp}$ just for convenience in our presentation.

The interaction terms with $g_{\perp}$ contain $\tau_{0}$ and $\tau_{3}$, while the interaction terms with $g_{\parallel}$ contain $\tau_{1}$ and $\tau_{2}$.
Because of our simplification $g_{\perp}=g'_{\perp}$, the Lagrangian~\eqref{eq:Lagrangian_general_Z2} is further invariant under the discrete ${\mathbb Z}_2$ symmetry, i.e., $\psi_{\L} \rightarrow \tau_{1}\psi_{\L}$ and $\psi_{\R} \rightarrow \tau_{1}\psi_{\R}$ simultaneously.
This symmetry, however, can be broken when the $g_{\perp} = g'_{\perp}$ condition is relaxed.

When the $\SU(2)_{\L} \times \SU(2)_{\R}$  emergent chiral symmetry is adopted,
the Lagrangian~\eqref{eq:Lagrangian_general_summary} becomes further simplified to
\begin{align}
   {\cal L}_{\mathrm{com}}^{\SU(2)_{\L} \times \SU(2)_{\R}}
&=
   \bar{\psi} i \gamma\partial \psi
 - 2g
   \bigl(
         (\bar{\psi}\tau_{0}\psi)^{2}
      + (\bar{\psi}\vect{\tau}\psi)^{2}
      + (\bar{\psi}i\gamma_{5}\tau_{0}\psi)^{2}
      + (\bar{\psi}i\gamma_{5}\vect{\tau}\psi)^{2}
   \bigr),
\label{eq:Lagrangian_general_SU2}
\end{align}
with $g=g_{\perp}=g_{\parallel}$.
Notice that the Lagrangian~\eqref{eq:Lagrangian_general_SU2} is the same as the Nambu--Jona-Lasinio model with the $\U(2)_{\L} \times \U(2)_{\R}$ (isospin) symmetry for light two-flavor quarks.~\footnote{Note that our $\psi$ carries no color.}
In the following, we mainly focus on the Lagrangian~\eqref{eq:Lagrangian_general_Z2} with 
the $\U(1)_{(1-2)\L} \times \U(1)_{(1-2)\R}$ emergent chiral symmetry.
The result for the Lagrangian~\eqref{eq:Lagrangian_general_SU2} with the $\SU(2)_{\L} \times \SU(2)_{\R}$ emergent chiral symmetry is easily obtained by setting $g_{\perp}=g_{\parallel}$.

\section{Mean field analyses} \label{sec:phase_diagram}

In this section, we construct the phase diagram of the emergent chiral superfluids in the mean-field approximation. The dynamical symmetry breakings and the topological properties of the order parameter space are discussed in detail. We also show the massless Dirac fermions near the Dirac points in momentum space at low-energy scales.

\subsection{Nambu-Gor'kov formalism}

As explained in the introduction, we are interested in the superfluids formed by the Cooper pairings.
The Cooper pairs are supplied not only by particle-particle pairings but also hole-hole pairings in the vicinity of the Fermi surface.
In order to describe the superfluids, we use the Nambu-Gor'kov formalism for the treatment of the particle-particle pairings and the hole-hole pairings on the same footing.
In order to adopt the Nambu-Gor'kov formalism at finite baryon density with chemical potential $\mu$, we rewrite the Lagrangian~\eqref{eq:Lagrangian_general_Z2} as
\begin{align}
   {\cal L}_{\mathrm{com}}^{\U(1)_{(1-2)\L} \times \U(1)_{(1-2)\R}\times {\mathbb Z}_2}
&=
   \frac{1}{2} \bar{\psi} \bigl( i \gamma\partial + \mu \gamma_{0} \bigr) \psi
+ \frac{1}{2} \bar{\psi}_{C} \bigl( i \gamma\partial - \mu \gamma_{0} \bigr) \psi_{C}
\nonumber \\ & 
 - g_{\perp}
   \sum_{a=0,3}
   (\bar{\psi}_{C}\gamma^{0}\gamma_{5}\tau_{a}\psi)^{\dag} (\bar{\psi}_{C}\gamma_{0}\gamma_{5}\tau_{a}\psi)
+ g_{\perp}
   \sum_{a=0,3}
   (\bar{\psi}_{C}\vect{\gamma}\gamma_{5}\tau_{a}\psi)^{\dag} (\bar{\psi}_{C}\vect{\gamma}\gamma_{5}\tau_{a}\psi)
   \nonumber \\ & 
 - g_{\parallel}
   (\bar{\psi}_{C}\gamma^{0}\gamma_{5}\tau_{1}\psi)^{\dag} (\bar{\psi}_{C}\gamma_{0}\gamma_{5}\tau_{1}\psi)
+ g_{\parallel}
   (\bar{\psi}_{C}\vect{\gamma}\gamma_{5}\tau_{1}\psi)^{\dag} (\bar{\psi}_{C}\vect{\gamma}\gamma_{5}\tau_{1}\psi)
   \nonumber \\ & 
 - g_{\parallel}
   (\bar{\psi}_{C}\gamma^{0}\tau_{2}\psi)^{\dag} (\bar{\psi}_{C}\gamma_{0}\tau_{2}\psi)
+ g_{\parallel}
   (\bar{\psi}_{C}\vect{\gamma}\tau_{2}\psi)^{\dag} (\bar{\psi}_{C}\vect{\gamma}\tau_{2}\psi),
\label{eq:Lagrangian_general_Z2_pairing}
\end{align}
by applying the Fierz transformations for the Dirac and color indices as summarized in Appendix~\ref{sec:Fierz_transformation}.
Here we have defined $\psi_{C}=C\gamma^{0}\psi^{\ast}$ with $C=i\gamma^{2}\gamma^{0}$
where $\gamma^{\mu}=(\gamma^{0},\vect{\gamma})$ is the Dirac matrix.
In Eq.~\eqref{eq:Lagrangian_general_Z2_pairing}, there are only the vector and axialvector channels and no other channels such as scalar and pseudoscalar channels.
As a result, we find
the possible quantum numbers of the pairings:
the emergent-chiral-symmetric (${\cal S}$) and vector ($J^{P}=1^{-}$) pairings
stemming from the bilinear form
$\bar{\psi}_{C}\vect{\gamma}\gamma_{5}\tau_{a}\psi$ ($a=0,1,3$)
and
the emergent-chiral-antisymmetric (${\cal A}$) and axialvector ($J^{P}=1^{+}$) pairings
from $\bar{\psi}_{C}\vect{\gamma}\tau_{2}\psi$.~\footnote{The condensates for the time components, $\bar{\psi}_{C}\gamma_{0}\gamma_{5}\tau_{a}\psi$ and $\bar{\psi}_{C}\gamma_{0}\tau_{2}\psi$, are not considered in the present study, as explained in the text.}
Here $J^{P}$ indicates the spin-parity as a usual notation.
We call the former the ${\cal S}_{a}1^{-}$ condensate and the latter the ${\cal A}1^{+}$ condensate, as summarized in Table~\ref{tbl:bilinear_form_quantum_number}.
Notice that
 the ${\cal S}_{a}1^{-}$ condensates are dependent on
 the directions in the $\SU(2)$ space ($a=0,1,3$).
We comment that the $1^{-}$ pairing is denoted by $^{3}P_{1}$ and the $1^{+}$ pairing by $^{3}S_{1}$-$^{3}D_{1}$ (mixing of $S$ and $D$ waves) when the nonrelativistic notation $^{2S+1}L_{J}$ with spin $S$ and angular momentum $L$ is adopted (cf.~Fig.~\ref{fig:240807_star_inside}).

\begin{table}[tp]
\begin{center}
\begin{tabular}{cc}
 \hline
 Bilinear form & quantum number (${\cal I}J^{P}$) \\ \hline
 $\bar{\psi}_{C}\vect{\gamma}\gamma_{5}\tau_{a}\psi$ & ${\cal S}_{a}1^{-}$ \\
 $\bar{\psi}_{C}\vect{\gamma}\tau_{2}\psi$ & ${\cal A}1^{+}$ \\
 \hline
\end{tabular}
\end{center}
\caption{Quantum numbers for the pairings in the parity-doubled baryons 
are summarized. The quantum numbers for each pairing are given by ${\cal I}={\cal S}_{a}$ ($a=0,1,3$) or ${\cal A}$ indicating the representation of the emergent-chiral-symmetric or -antisymmetric representation and $J^{P}=1^{-}$ or $1^{+}$ for the spin-parity.}
\label{tbl:bilinear_form_quantum_number}
\end{table}%

We classify the possible patterns of the
stable pairings 
according to the signs of the coupling constants in the Lagrangian~\eqref{eq:Lagrangian_general_Z2_pairing}: (i) $g_{\perp}>0$ and $g_{\parallel}>0$, (ii) $g_{\perp}>0$ and $g_{\parallel}<0$, (iii) $g_{\perp}<0$ and $g_{\parallel}>0$, and (iv) $g_{\perp}<0$ and $g_{\parallel}<0$.
When either $g_{\perp}$ or $g_{\parallel}$ is negative, i.e., in the cases (ii), (iii), and (iv), we find 
that the time components of the interaction terms in Eq.~\eqref{eq:Lagrangian_general_Z2_pairing} shift the chemical potential, and they play no important role for the pairing formations.
Therefore, in the following, we concentrate on case (i).

In case (i),
we may
neglect the repulsive parts in the interaction terms in Eq.~\eqref{eq:Lagrangian_general_Z2_pairing} and leave the attractive parts only. Then, we simplify the Lagrangian as
\begin{align}
   {\cal L}_{\mathrm{\com}}^{\U(1)_{(1-2)\L} \times \U(1)_{(1-2)\R}\times {\mathbb Z}_2}
&\approx
   \frac{1}{2} \bar{\psi} \bigl( i \gamma\partial + \mu \gamma_{0} \bigr) \psi
+ \frac{1}{2} \bar{\psi}_{C} \bigl( i \gamma\partial - \mu \gamma_{0} \bigr) \psi_{C}
\nonumber \\ & 
+ g_{\perp}
   (\bar{\psi}_{C}\vect{\gamma}\gamma_{5}\tau_{0}\psi)^{\dag} (\bar{\psi}_{C}\vect{\gamma}\gamma_{5}\tau_{0}\psi)
+ g_{\parallel}
   (\bar{\psi}_{C}\vect{\gamma}\gamma_{5}\tau_{1}\psi)^{\dag} (\bar{\psi}_{C}\vect{\gamma}\gamma_{5}\tau_{1}\psi)
+ g_{\perp}
   (\bar{\psi}_{C}\vect{\gamma}\gamma_{5}\tau_{3}\psi)^{\dag} (\bar{\psi}_{C}\vect{\gamma}\gamma_{5}\tau_{3}\psi)
   \nonumber \\ & 
+ g_{\parallel}
   (\bar{\psi}_{C}\vect{\gamma}\tau_{2}\psi)^{\dag} (\bar{\psi}_{C}\vect{\gamma}\tau_{2}\psi).
\label{eq:Z2PF_g_negative_negative_Lagrangian}
\end{align}
Based on this Lagrangian,
we construct the thermodynamic potential by the Nambu-Gor'kov formalism in which particle-particle (hole-hole) pairings are described.
Following the usual prescription, we introduce the auxiliary boson (complex) fields 
\begin{eqnarray}
\vect{\Delta}_{a}=(\Delta_{ai}) 
\sim \bar{\psi}_{C}\vect{\gamma}\gamma_{5}\tau_{a}\psi
=
\bar{\psi}_{\L C}\vect{\gamma}\gamma_{5}\tau_{a}\psi_{\R}+\bar{\psi}_{\R C}\vect{\gamma}\gamma_{5}\tau_{a}\psi_{\L},
\end{eqnarray}
and
\begin{eqnarray}
\vect{\delta}=(\delta_{i}) 
\sim \bar{\psi}_{C}\vect{\gamma}\tau_{2}\psi
= \bar{\psi}_{\L C}\vect{\gamma}\tau_{2}\psi_{\R} + \bar{\psi}_{\R C}\vect{\gamma}\tau_{2}\psi_{\L},
\end{eqnarray}
with $a=0,1,3$ for the $\SU(2)$ indices and $i=1,2,3$ for 
spatial vector indices.
Note that $\vect{\Delta}_{a}$ and $\vect{\delta}$ are proportional to the expectation values of $\bar{\psi}_{C}\vect{\gamma}\gamma_{5}\tau_{a}\psi$ and $\bar{\psi}_{C}\vect{\gamma}\tau_{2}\psi$, respectively.
Calculating the generating functionals for the Lagrangian~\eqref{eq:Z2PF_g_negative_negative_Lagrangian}
in the one-loop approximation,
 we obtain the thermodynamic potential given by
\begin{align}
   W
&=
 - \Tr
   \ln
   G^{-1}(\{\vect{\Delta}_{a}\},\vect{\delta})
+ \frac{1}{4}
   \int \dr^{4}x
   \biggl(
         \frac{1}{g_{\perp}} \vect{\Delta}_{0}^{\dag} \vect{\Delta}_{0}
      + \frac{1}{g_{\parallel}} \vect{\Delta}_{1}^{\dag} \vect{\Delta}_{1}
      + \frac{1}{g_{\perp}} \vect{\Delta}_{3}^{\dag} \vect{\Delta}_{3}
      + \frac{1}{g_{\parallel}} \vect{\delta}^{\dag} \vect{\delta}
   \biggr),
\label{eq:general_Z2_negative_negative_generating_functional_connected}
\end{align}
with the inverse of the dressed propagator
\begin{align}
   G^{-1}(\{\vect{\Delta}_{a}\},\vect{\delta})
=
   \left(
   \begin{array}{cc}
      \gamma\partial - \mu \gamma^{4} &
      \sum_{a} \vect{\gamma}\gamma^{5}\tau_{a} \vect{\Delta}_{a}
      + \vect{\gamma}\tau_{2} \vect{\delta} \\ 
       - \sum_{a} \vect{\gamma}\gamma^{5}\tau_{a} \vect{\Delta}_{a}^{\dag}
       - \vect{\gamma}\tau_{2} \vect{\delta}^{\dag} &
      \gamma\partial + \mu \gamma^{4}
   \end{array}
   \right),
\label{eq:general_Z2_propagator}
\end{align}
in the four-dimensional Euclidean space for time and space.
Here $\Tr$ indicates the trace over all the points in the Euclidean space.
In Eq.~\eqref{eq:general_Z2_negative_negative_generating_functional_connected}, the most stable state is obtained by the variational calculation for $W$ with respect to $\vect{\Delta}_{a}$ ($a=0,1,3$) and $\vect{\delta}$.
We define the density $w$ by $\displaystyle W = \int \dr^{4} \, w$ as the useful quantity in the following analysis.

\subsection{Phase diagram}

The pairing formation in the superfluids induce the nonzero condensate for each  $\vect{\Delta}_{a}$ and $\vect{\delta}$ in Eq.~\eqref{eq:general_Z2_negative_negative_generating_functional_connected}.
As a simple situation, we suppose that the condensates $\vect{\Delta}_{a}$ and $\vect{\delta}$ are 
uniformly directed into the $z$ direction in the real space as constant vectors: $\vect{\Delta}_{a}=(0,0,\Delta_{a})$ for the ${\cal S}_{a}1^{-}$ condensate ($a=0,1,3$) and $\vect{\delta}=(0,0,\delta)$ for the ${\cal A}1^{+}$ condensate.
Here, $\Delta_{a}$ and $\delta$ are assumed to be constant values due to the uniformity.
With this assumption, we obtain the single-particle energy for the (gapped) quasiparticles, stemming from the energy eigenvalues in the dressed propagator, $G(\{\vect{\Delta}_{a}\},\vect{\delta})$, in Eq.~\eqref{eq:general_Z2_propagator}.

\begin{figure}
\includegraphics[keepaspectratio, scale=0.45]{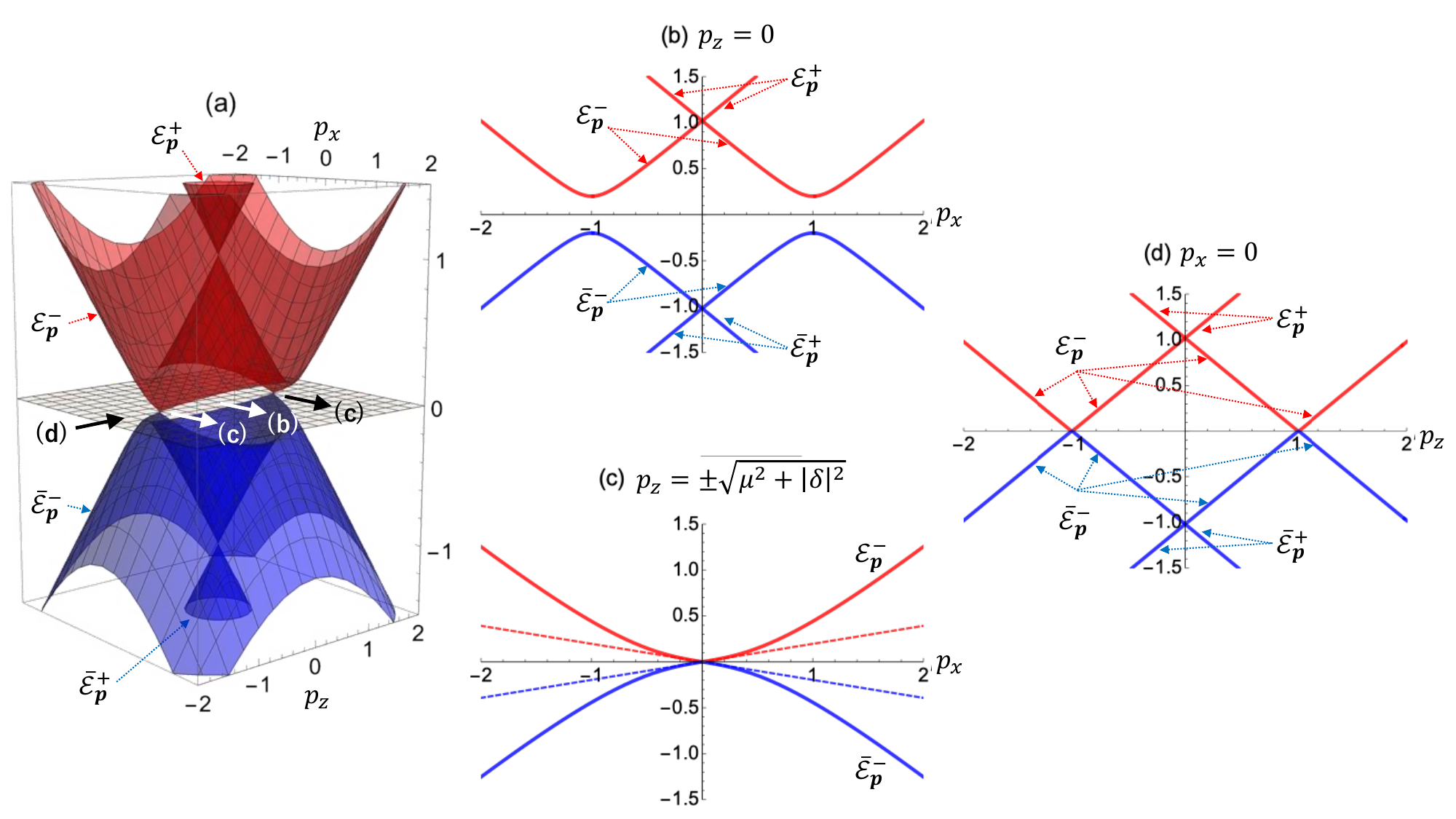}%
\caption{
Single-particle energies, ${\cal E}_{\vect{p}}^{+}$, ${\cal E}_{\vect{p}}^{-}$, $\bar{\cal E}_{\vect{p}}^{+}$ and $\bar{\cal E}_{\vect{p}}^{-}$, for the ${\cal A}1^{+}$ condensate with $\vect{\delta}=(0,0,\delta)$ are plotted.
The values of the parameters are $\mu=1$ and $|\delta|=0.2$.
The red surfaces (lines) at positive energy indicate ${\cal E}_{\vect{p}}^{+}$ and ${\cal E}_{\vect{p}}^{-}$, and the blue surfaces (lines) at negative energy indicate $\bar{\cal E}_{\vect{p}}^{+}$ and $\bar{\cal E}_{\vect{p}}^{-}$.
(a) The energies are plotted on the $p_{x}$-$p_{z}$ plane with $p_{y}=0$.
(b) The energies are plotted on the $p_{x}$ axis with $p_{z}=0$.
(c) The energies for $p_{z}=\pm\sqrt{\mu^{2}+|\delta|^{2}}$ are plotted on the $p_{x}$ axis with $p_{y}=0$.
The dashed lines represent the approximate linear solutions in Eq.~\eqref{eq:SU2PF_hamiltonian_ansatz_Dirac_point_eff_ham_sol}.
(d) The energies are plot on the $p_{z}$ axis.
Notice that, in (a), the cusps between the red and blue surfaces indicate the Dirac points at $\vect{p}_{\pm}^{\ast}$ given in Eq.~\eqref{eq:SU2PF_Dirac_point}. Notice also that, in (c), the red and blue dashed lines have the cusps (the discontinuity in the derivatives) at the Dirac points $p_{x}=0$ and $p_{z}=\pm\sqrt{\mu^{2}+|\delta|^{2}}$, respectively, as indicated in Eq.~\eqref{eq:SU2PF_hamiltonian_ansatz_Dirac_point_eff_ham_sol}.
\label{fig:231230}}
\end{figure}

 We show the energy-momentum dispersion relations for each the ${\cal S}_{a}1^{-}$ condensate ($a=0,1,3$) and the ${\cal A}1^{+}$ condensate.
For the ${\cal S}_{a}1^{-}$ condensate, we set $\vect{\Delta}_{a}=(0,0,\Delta_{a})$ and $\vect{\delta}=\vect{0}$.
Then, we obtain the single-particle energy eigenvalues parametrized by $\Delta_{a}$ as
\begin{align}
   {\cal E}_{\vect{p}}^{\pm}(\Delta_{a})
&=
   \sqrt{\vect{p}^{2}+\mu^{2}+|\Delta_{a}|^{2} \pm 2\sqrt{\vect{p}^{2}\mu^{2}+p_{z}^{2}|\Delta_{a}|^{2}}},
\nonumber \\ 
   \bar{\cal E}_{\vect{p}}^{\pm}(\Delta_{a})
&=
 - \sqrt{\vect{p}^{2}+\mu^{2}+|\Delta_{a}|^{2} \pm 2\sqrt{\vect{p}^{2}\mu^{2}+p_{z}^{2}|\Delta_{a}|^{2}}},
\label{eq:energy_momentum_dispersion_Z2_negarive_negative_S}
\end{align}
with the three-dimensional momentum $\vect{p}=(p_{x},p_{y},p_{z})$.
For the quasiparticles with gap $\Delta_{a}$, there are four degenerate solutions for ${\cal E}_{\vect{p}}^{\pm}(\Delta_{a})$ in each $\pm$. It is the same also for $\bar{\cal E}_{\vect{p}}^{\pm}(\Delta_{a})$.
For the ${\cal A}1^{+}$ condensate, we set $\vect{\Delta}_{a}=\vect{0}$ and $\vect{\delta}=(0,0,\delta)$.
Then, we obtain similarly the single-particle energy eigenvalues parametrized by $\delta_{a}$ as
\begin{align}
   {\cal E}_{\vect{p}}^{\pm}(\delta)
&=
   \sqrt{\vect{p}^{2}+\mu^{2}+|\delta|^{2} \pm 2\sqrt{\vect{p}^{2}\mu^{2}+p_{z}^{2}|\delta|^{2}}},
\nonumber \\ 
   \bar{\cal E}_{\vect{p}}^{\pm}(\delta)
&=
 - \sqrt{\vect{p}^{2}+\mu^{2}+|\delta|^{2} \pm 2\sqrt{\vect{p}^{2}\mu^{2}+p_{z}^{2}|\delta|^{2}}}.
\label{eq:energy_momentum_dispersion_Z2_negarive_negative_A}
\end{align}
For the quasiparticles with gap $\delta$, there are four degenerate solutions for ${\cal E}_{\vect{p}}^{\pm}(\delta)$ in each $\pm$. It is the same also for $\bar{\cal E}_{\vect{p}}^{\pm}(\delta)$.
Notice the anisotropy in the momentum space due to the choice of the $z$ axis in the vector condensates.
As a numerical example,
we plot ${\cal E}_{\vect{p}}^{\pm}(\delta)$ and $\bar{\cal E}_{\vect{p}}^{\pm}(\delta)$
for the ${\cal A}1^{+}$ condensate in Fig.~\ref{fig:231230}.

\begin{figure}
\includegraphics[keepaspectratio, scale=0.3]{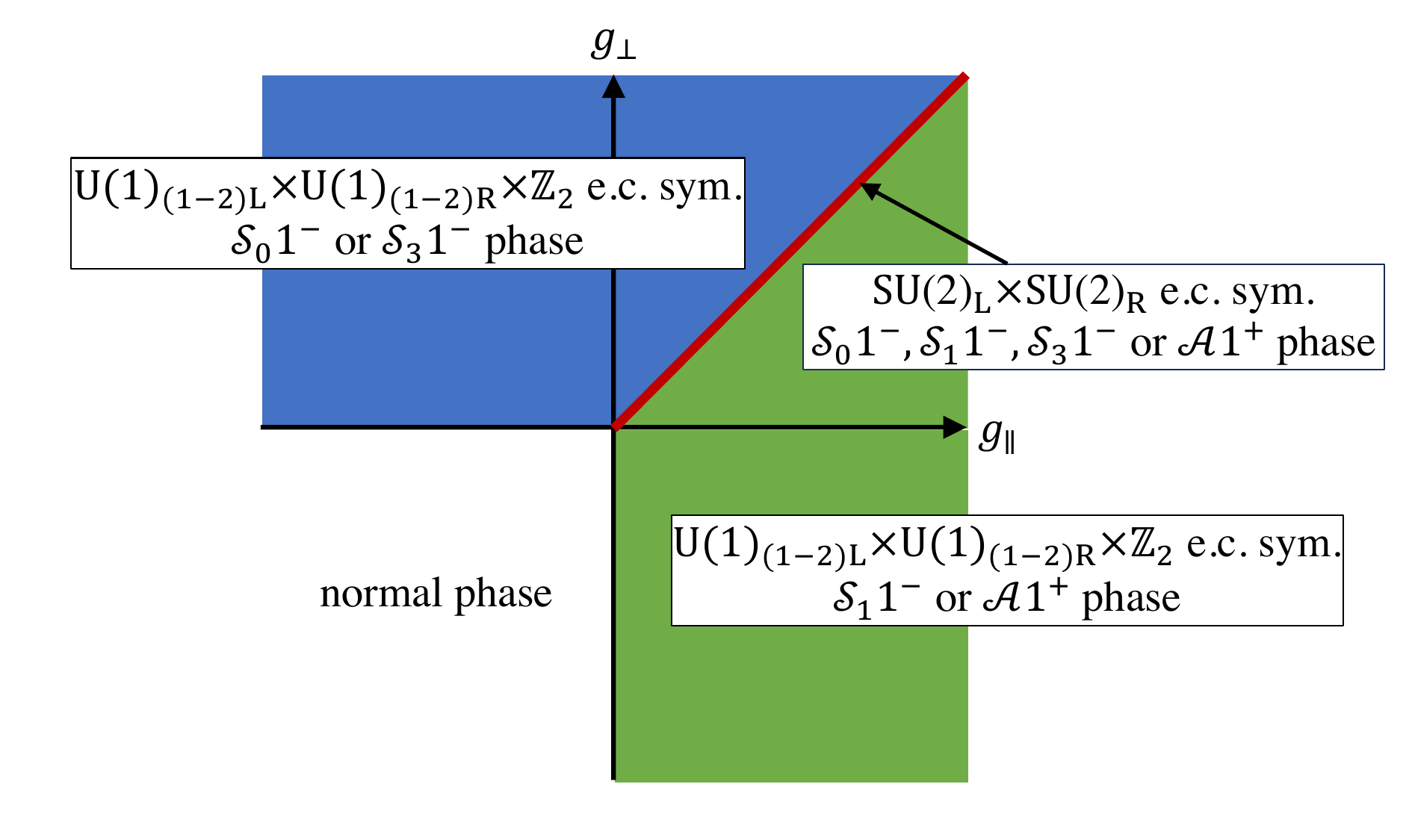} %
\caption{
Schematic figure of the phase diagram is shown for the emergent chiral superfluids with the 
$\U(1)_{(1-2)\L} \times \U(1)_{(1-2)\R} \times {\mathbb Z}_{2}$ emergent chiral symmetry [i.e., the $\frac{\U(1)_{\L} \times \U(1)_{(1-2)\L}}{{\mathbb Z}_{2\L}^\prime}
 \times 
 \frac{\U(1)_{\R} \times \U(1)_{(1-2)\R}}{{\mathbb Z}_{2\R}^\prime} \times {\mathbb Z}_{2}$ full symmetry] and the $\SU(2)_{\L} \times \SU(2)_{\R}$ emergent chiral symmetry [i.e., the 
$\U(2)_{\L} \times \U(2)_{\R}$ full symmetry].
Here e.c. sym. denotes the emergent chiral symmetry.
In the 
$\U(1)_{(1-2)\L} \times \U(1)_{(1-2)\R}$ emergent chiral symmetry, the ${\cal S}_{0}1^{-}$ or ${\cal S}_{3}1^{-}$ condensate exists in the blue region ($g_{\perp} > g_{\parallel}$ and $g_{\perp}>0$), and the ${\cal S}_{1}1^{-}$ or ${\cal A}1^{+}$ condensate exists in the green region ($g_{\perp} < g_{\parallel}$ and $g_{\parallel}>0$).
In the $\SU(2)_{\L} \times \SU(2)_{\R}$ emergent chiral symmetry, the ${\cal S}_{0}1^{-}$, ${\cal S}_{1}1^{-}$, ${\cal S}_{3}1^{-}$, and ${\cal A}1^{+}$ condensates are degenerate on the red semiline ($g_{\perp} = g_{\parallel}>0$).}
\label{fig:240902_phase_diagram}
\end{figure}

With the help of Eqs.~\eqref{eq:energy_momentum_dispersion_Z2_negarive_negative_S} and \eqref{eq:energy_momentum_dispersion_Z2_negarive_negative_A}, we obtain the thermodynamic potential densities for each of the ${\cal S}_{a}1^{-}$ and ${\cal A}1^{+}$ condensates.
We denote the potential densities for the ${\cal S}_{0}1^{-}$ and ${\cal S}_{3}1^{-}$ condensates by $w_{\perp}(\Delta_{0})$ and $w_{\perp}(\Delta_{3})$, respectively, where the function $w_{\perp}(\xi)$ is given by
\begin{align}
   w_{\perp}(\xi)
&=
 - \int \frac{\dr^{3}\vect{p}}{(2\pi)^{3}}
   \frac{2}{\beta}
   \ln
   \Biggl(
         \cosh^{4}\biggl(\frac{1}{2}\beta{\cal E}_{\vect{p}}^{+}(\xi)\biggr)
         \cosh^{4}\biggl(\frac{1}{2}\beta{\cal E}_{\vect{p}}^{-}(\xi)\biggr)
   \Biggr)
+ \frac{1}{4g_{\perp}}
   |\xi|^{2},
\label{eq:general_Z2PF_negative_negarive_generating_functional_connected_density_ansatz_03}
\end{align}
with $\xi=\Delta_{0}$ or $\Delta_{3}$.
Here $\beta=1/T$ the inverse of temperature.
Similarly, we denote the potential densities for the ${\cal S}_{1}1^{-}$ and ${\cal A}1^{+}$ condensates by $w_{\parallel}(\Delta_{1})$ and $w_{\parallel}(\delta)$, respectively, where the function $w_{\parallel}(\xi)$ is given by
\begin{align}
   w_{\parallel}(\xi)
&=
 - \int \frac{\dr^{3}\vect{p}}{(2\pi)^{3}}
   \frac{2}{\beta}
   \ln
   \Biggl(
         \cosh^{4}\biggl(\frac{1}{2}\beta{\cal E}_{\vect{p}}^{+}(\xi)\biggr)
         \cosh^{4}\biggl(\frac{1}{2}\beta{\cal E}_{\vect{p}}^{-}(\xi)\biggr)
   \Biggr)
+ \frac{1}{4g_{\parallel}}
   |\xi|^{2},
\label{eq:general_Z2PF_negative_negarive_generating_functional_connected_density_ansatz_12}
\end{align}
with $\xi = \Delta_{1}$ or $\delta$.
The most stable states are obtained by minimizing each of $w_{\perp}(\Delta_{0})$, $w_{\perp}(\Delta_{3})$, $w_{\parallel}(\Delta_{1})$ and $w_{\parallel}(\delta)$ with respect to $\Delta_{0}$, $\Delta_{3}$, $\Delta_{1}$ and $\delta$, respectively.~\footnote{The possibility of the coexistence of different condensates, e.g., the coexistence of the ${\cal S}_{a}1^{-}$ and ${\cal A}1^{+}$ condensates, is not considered in the present study.}
Among them, the most stable states at zero temperature are determined by the relations among the coupling constants $g_{\perp}$ and $g_{\parallel}$
\begin{align}
   g_{\perp} > g_{\parallel} \,\, \mathrm{and} \,\, g_{\perp}>0 &: \,\, {\cal S}_{0}1^{-} \,\,\mathrm{or}\,\, {\cal S}_{3}1^{-} \,\, \mathrm{condensate},
   \nonumber \\ 
   g_{\perp} < g_{\parallel} \,\, \mathrm{and} \,\, g_{\parallel}>0 &: \,\, {\cal S}_{1}1^{-} \,\,\mathrm{or}\,\, {\cal A}1^{+} \,\, \mathrm{condensate}.
\label{sec:Z2PF_g_negative_negative_cond}
\end{align}
In the above equation, the ${\cal S}_{0}1^{-}$ and ${\cal S}_{3}1^{-}$ condensates are degenerate, i.e., $w_{\perp}(\Delta_{0})=w_{\perp}(\Delta_{3})$, and the ${\cal S}_{1}1^{-}$ and ${\cal A}1^{+}$ are also degenerate, i.e., $w_{\parallel}(\Delta_{1})=w_{\parallel}(\delta)$ (see Table~\ref{tbl:Z2_SU2PF_condensate}).
This degeneracy stems from the  
$\U(1)_{(1-2)\L} \times \U(1)_{(1-2)\R}$ emergent chiral 
symmetry.
When the emergent chiral symmetry is 
enlarged
to the 
$\SU(2)_{\L} \times \SU(2)_{\R}$
symmetry,
 we set $g_{\perp}=g_{\parallel}>0$ and find that any of the four condensates is realized and takes the same value among them: $w_{\perp}(\Delta_{0})=w_{\perp}(\Delta_{3})=w_{\parallel}(\Delta_{1})=w_{\parallel}(\delta)$.
As a summary, we obtain the schematic picture of the phase diagram  at zero temperature shown in Fig.~\ref{fig:240902_phase_diagram}.

\begin{table}[tp]
\begin{center}
\begin{tabular}{ccc}
 \hline
 Emergent chiral symmetry & Coupling relation & Condensate \\
 \hline
 $\U(1)_{(1-2)\L} \times \U(1)_{(1-2)\R} 
 \times {\mathbb Z}_2$
 & $g_{\perp}>g_{\parallel}$ and $g_{\perp}>0$ & ${\cal S}_{0}1^{-}$, ${\cal S}_{3}1^{-}$ \\
  & $g_{\perp}<g_{\parallel}$ and $g_{\parallel}>0$ & ${\cal S}_{1}1^{-}$, ${\cal A}1^{+}$ \\
 $\SU(2)_{\L} \times \SU(2)_{\R}$ & $g_{\perp}=g_{\parallel}>0$ & ${\cal S}_{0}1^{-}$, ${\cal S}_{1}1^{-}$, ${\cal S}_{3}1^{-}$, ${\cal A}1^{+}$ \\
 \hline
\end{tabular}
\end{center}
\caption{Condensates in the ground state are summarized for different emergent chiral symmetries and the coupling relations.}
\label{tbl:Z2_SU2PF_condensate}
\end{table}%

So far, we have considered the $\U(1)_{(1-2)\L} \times \U(1)_{(1-2)\R} \times {\mathbb Z}_{2}$ emergent chiral symmetry by setting $g_{\perp}=g'_{\perp}$.
We may consider the original $\U(1)_{(1-2)\L} \times \U(1)_{(1-2)\R}$ emergent chiral symmetry by allowing $g_{\perp}$ and $g'_{\perp}$ to be different more generally.
The condition $g_{\perp} \neq g'_{\perp}$ would lead to resolving the degeneracy shown in the phase diagram in Fig.~\ref{fig:240902_phase_diagram};
either the ${\cal S}_{0}1^{-}$ condensate or the ${\cal S}_{3}1^{-}$ condensate will be realized, and either the  ${\cal S}_{1}1^{-}$ condensate or the ${\cal A}1^{+}$ condensate will be realized.
Similarly, on the semiline, only two degenerate condensates among the ${\cal S}_{0}1^{-}$, ${\cal S}_{1}1^{-}$, ${\cal S}_{3}1^{-}$ and ${\cal A}1^{+}$ condensates will be realized.
In any case, because there exist emergent chiral superfluids in the ground state, our conclusion is not affected in essence.

\subsection{Dynamical symmetry breakings and their topology}\label{sec:DSB}

There is a particular pattern of symmetry breaking in the ${\cal S}_{0}1^{-}$, ${\cal S}_{3}1^{-}$, ${\cal S}_{1}1^{-}$ and ${\cal A}1^{+}$ condensates.
In Eq.~\eqref{eq:emergent-chiral3},
we have considered the 
$\SU(2)_{\L} \times \SU(2)_{\R}$
as the extension of emergent chiral symmetry 
 $\U(1)_{(1-2)\L} \times \U(1)_{(1-2)\R}$.  
Including the $\SO(3)_{\rm S}$ 
rotational symmetry in the three-dimensional real space, the total symmetries are given by
\begin{align}
 &
 \frac{\U(1)_{\L} \times 
       \U(1)_{(1-2)\L} }{
   {\mathbb Z}_{2\L}^\prime}
   \times 
   \frac{\U(1)_{\R} \times \U(1)_{(1-2)\R}}{
   {\mathbb Z}_{2\R}^\prime}
   \times \SO(3)_{\rm S} \nonumber \\
& \subset
   \U(2)_{\L} \times \U(2)_{\R} \times \SO(3)_{\rm S}.
\end{align}
When the ${\cal S}_{a}1^{-}$ ($a=0,1,3$) or ${\cal A}1^{+}$ condensation occurs, 
each symmetry is 
dynamically broken as 
\begin{align}
& \frac{\U(1)_{\L} \times 
       \U(1)_{(1-2)\L} }{
   {\mathbb Z}_{2\L}^\prime}
   \times 
   \frac{\U(1)_{\R} \times \U(1)_{(1-2)\R}}{
   {\mathbb Z}_{2\R}^\prime} \times {\mathbb Z}_2
   \times \SO(3)_{\rm S} \nonumber\\
  & \quad\quad\longrightarrow
  \U(1)_{\L-\R} \times
 \U(1)_{(1-2)(\L+\R)} 
   \times {\mathbb Z}_2 \times
   [\SO(2)_{\rm S} \rtimes 
   {\mathbb Z}_{2({\rm L+R+S})}],
\label{eq:symmetry_breaking_U(1)}
\\
&   \U(2)_{\L} \times \U(2)_{\R} \times \SO(3)_{\rm S}
\longrightarrow
   \U(1)_{\L-\R} \times \SU(2)_{\L+\R} 
   \times 
   [\SO(2)_{\rm S} \rtimes 
   {\mathbb Z}_{2({\rm L+R+S})}],
\label{eq:symmetry_breaking_SU2}
\end{align}
for the $\U(1)_{(1-2)\L} \times \U(1)_{(1-2)\R}$ 
and $\SU(2)_{\L} \times \SU(2)_{\R}$ emergent chiral symmetries, 
respectively.
In all the cases, 
the baryon number symmetry $\U(1)_{\L+\R}$ is spontaneously 
broken, resulting in superfluidity.
The chiral (axial) symmetry $\U(1)_{\L-\R}$ is unbroken.
In contrast,
the 
emergent chiral symmetries,
$\U(1)_{(1-2)\L} \times 
\U(1)_{(1-2)\R}$ 
and $\SU(2)_{\L}
\times \SU(2)_{\R}$,
are spontaneously broken to 
$\U(1)_{(1-2)(\L+\R)}$ 
and 
$\SU(2)_{\L+\R}$,
i.e., vector-like emergent chiral symmetries
interlocked by
the left and right fermions, respectively.
This implies
the emergent chiral symmetry breaking.
 As for the spatial rotation, 
 the $\SO(3)_{\rm S}$ rotational symmetry 
 is broken to the  
 $\SO(2)_{\rm S}$ symmetry,  
 a rotation around the $z$-axis  
keeping the vector-type condensates such as $\vect{\Delta}_{a}=(0,0,\Delta_{a})$ for the ${\cal S}_{a}1^{-}$ condensate ($a=0,1,3$) and $\vect{\delta}=(0,0,\delta)$ for the ${\cal A}1^{+}$ condensate.
The 
 ${\mathbb Z}_{2({\rm L+R+S})}$
 is defined as 
the action of 
a $\pi$ rotation 
around the $x$ or $y$ axis 
represented by 
 $\diag (1,-1,-1)$ or 
 $\diag (-1,1,-1)
 \in \SO(3)_{\rm S}$
 together with a 
  $\pi$ phase rotation of the baryon symmetry 
  $\vect{\Delta}_{a} \to 
  e^{i\pi} \vect{\Delta}_{a},
  \vect{\delta} \to    e^{i\pi} \vect{\delta}  $
 (or $\psi_{\L} \to e^{i\pi/2}\psi_{\L}$,
 $\psi_{\R} \to e^{i\pi/2}\psi_{\R}$ in terms of fermions degrees of freedom).~\footnote{
This breaking pattern is the same with the polar phase of spin-1 
spinor Bose-Einstein condensates~\cite{Kawaguchi:2012ii}.
\label{foot:BEC}
}
Therefore, in such a way, the ${\cal S}_{a}1^{-}$ ($a=0,1,3$) or  ${\cal A}1^{+}$ condensate induces a simultaneous dynamical breaking of the baryon number,  
emergent chiral and 
rotational symmetries.

We leave some comments on the
symmetry breakings
in Eqs.~\eqref{eq:symmetry_breaking_U(1)}
and \eqref{eq:symmetry_breaking_SU2}.
They indicate the new patterns of symmetry breaking at finite density.
We may expect naively that the emergent chiral symmetry is usually restored
at sufficiently high density.
Indeed, the starting point of our discussion is based on the idea that the nucleon $N$ ($1/2^{+}$) and the excited nucleon $N^{\ast}(1535)$ ($1/2^{-}$) are degenerate at high density as 
the
parity-doubled baryons.
Contrary to this, 
the present study shows that the emergent chiral symmetry is dynamically broken by the pairings in the superfluids.
It is also interesting that the vector-type condensate is realized in the superfluids.
This is again different from the normal case that the scalar-type condensate is  favored usually.
Especially, the vector-type condensate cannot be seen in vacuum, because the realization of vector-type condensate is forbidden due to the Vafa-Witten theorem.~\footnote{The Vafa-Witten theorem indicates that that vector-like global symmetries cannot be spontaneously broken in the ground state~\cite{Vafa:1983tf,Vafa:1984xg}.}
Thus, the emergent chiral superfluids exhibit unique features which are not seen in normal superfluids.

Now, let us discuss 
massless bosonic excitations associated
with the dynamical symmetry breakings,
i.e., NG modes.
The order parameter spaces (OPS) 
for the dynamical symmetry breakings, parametrized by NG modes, can be written as 
\begin{align}
 M = M_{\rm emergent\; chiral} \times M_{\rm{baryon,space}}.
 \label{eq:OPS}
\end{align}
Here $M_{\rm emergent\; chiral}$ is 
an OPS  
associated with the emergent chiral symmetry breakings, given by
\begin{align}
\renewcommand{\arraystretch}{2}
M_{\rm emergent\; chiral} 
\simeq 
\frac{ G_{\L} \times G_{\R} }
   { G_{\L+\R}}
= 
\left\{
\begin{array}{ccc}
 \displaystyle{
 \frac{ \U(1)_{(1-2)\L} \times \U(1)_{(1-2)\R} }
   { \U(1)_{(1-2)(\L+\R)}}
   }
 &   \simeq &
\U(1)_{(1-2)(\L-\R)}
\\
   \displaystyle{
   \frac{\SU(2)_{\L} \times \SU(2)_{\R}}
   {\SU(2)_{\L+\R}}
   }
 &  \simeq &
 \SU(2)_{\L-\R},
 \end{array}
 \right.
\label{eq:OPS2}
\renewcommand{\arraystretch}{1}
\end{align}
for the $G_{\L} \times G_{\R}=\U(1)_{(1-2)\L} \times \U(1)_{(1-2)\R}$
and $\SU(2)_{\L} \times \SU(2)_{\R}$ emergent chiral symmetries, 
respectively. 
The NG modes corresponding to these breakings are all of type-I with linear dispersion relations~\cite{Watanabe:2012hr,Hidaka:2012ym,Takahashi:2014vua}.
They may be called ``emergent pions'' 
and 
consist of six quarks, 
with the baryon number two. 
These massless bosons can obtain masses to become pseudo-NG bosons~\cite{Weinberg:1972fn} (like ordinary pions) when we add terms explicitly breaking the emergent chiral symmetries to the original Lagrangian.
The naive and mirror assignments considered in the literature are such cases.

On the other hand, 
$M_{\rm{baryon,space}}$ in Eq.~\eqref{eq:OPS} is 
an OPS associated with the dynamical breaking of the baryon number symmetry 
and spatial rotational symmetry, 
given by 
\begin{align}
 M_{\rm{baryon,space}} = 
  \frac{ \U(1)_{\L} \times \U(1)_{\R} 
   \times 
   \SO(3)_{\rm S} }
   { \U(1)_{\L-\R} \times 
   \SO(2)_{\rm S} \rtimes 
   {\mathbb Z}_{2({\rm L+R+S})}} 
  \simeq \frac{\U(1)_{\L+\R} \times S^2_{\rm S} }
   {{\mathbb Z}_{2({\rm L+R+S})}} ,
   \label{eq:OPS-bs}
\end{align}
where $S^2_{\rm S} \simeq  \SO(3)_{\rm S}/\SO(2)_{\rm S}$.
This OPS, $M_{\rm baryon,space}$, is the same with 
that of the polar phase of the spin-1 Bose-Einstein condensate (BEC); 
see footnote~\ref{foot:BEC}.
As mentioned, 
the spontaneous breaking of 
the baryon number symmetry
$\U(1)_{\L+\R}$ is responsible for superfluidity. 
NG modes corresponding to this OPS are 
a superfluid phonon as a type-I NG mode with a linear dispersion relation for the $\U(1)_{\L+\R}$ breaking 
and magnons as a type-II NG mode with a 
quadratic dispersion relation for
the $\SO(3)_{\rm S}$ symmetry breaking~\cite{Takahashi:2014vua}.
Unlike the NG bosons (emergent pions) for the emergent chiral symmetry breakings, these modes are exactly massless since the baryon and rotational symmetries are exact.

\begin{table}[tp]
\begin{center}
\begin{tabular}{c|c|cccc}
 \hline 
 $G_{\L} \times G_{\R}$ & $M_{\rm{e.c.}}$ & 
 $\pi_0(M_{\rm{e.c.}})$ & $\pi_1(M_{\rm{e.c.}})$ & $\pi_2(M_{\rm{e.c.}})$ & $\pi_3(M_{\rm{e.c.}})$\\
 \hline
 $\U(1)_{(1-2)\L} \times \U(1)_{(1-2)\R}$ & 
 $\U(1)_{(1-2)(\L-\R)}$
 & $0$ & ${\mathbb Z}$ & $0$ & $0$ \\
 $\SU(2)_{\L} \times \SU(2)_{\R}$ 
 & $\SU(2)_{\L-\R}$ 
 & $0$ & $0$ & $0$ & ${\mathbb Z}$ \\
 \hline
\end{tabular}
\end{center}
\caption{
Lower dimensional homotopy groups 
from $\pi_0$ to $\pi_3$ 
are summarized 
for the emergent chiral symmetry breakings.
$M_{\rm e.c.}$ denotes 
the OPS $M_{\rm emergent\; chiral}$ in Eq.~\eqref{eq:OPS2} for these symmetry breakings.
\label{tbl:homotopy}}
\end{table}%

Lastly,
let us discuss topology and topological solitons associated with the dynamical symmetry breakings.
The lower-dimensional 
homotopy groups 
$\pi_0$ to $\pi_3$
of $M_{\rm emergent\; chiral}$ are 
summarized in Table~\ref{tbl:homotopy}.
Here $0$ indicates a trivial group 
admitting no topological solitons.
The $\U(1)_{\L} \times \U(1)_{\R}$
emergent chiral symmetry breaking admits axial strings supported by $\pi_1$.~\footnote{
In the presence of explicit breaking terms, this axial string is attached by domain walls~\cite{Eto:2013hoa,Eto:2013bxa}.
}
The $\SU(2)_{\L} \times \SU(2)_{\R}$ emergent chiral symmetry breaking 
admits Skyrmions supported by $\pi_3$.

On the other hand, 
the lower-dimensional 
homotopy groups of
$M_{\rm{baryon,space}}$ are
\begin{align}
\pi_0 (M_{\rm{baryon,space}}) 
\simeq 0, 
\quad 
\pi_1 (M_{\rm{baryon,space}}) 
\simeq {\mathbb Z}, 
\quad 
\pi_2 (M_{\rm{baryon,space}}) 
\simeq {\mathbb Z}, 
\quad
\pi_3 (M_{\rm{baryon,space}}) 
\simeq {\mathbb Z}.
\end{align}
These admit various topological solitons~\cite{Kawaguchi:2012ii}:  
superfluid vortices called 
singly quantized vortices and half-quantized Alice strings~\cite{Leonhardt:2000km, Chatterjee:2017jsi}
 supported by $\pi_1$, 
lump-strings (of codimension two) 
and monopoles (of codimension three)~\cite{Ruostekoski:2003qx} 
supported
by 
$\pi_2$, 
and Hopfions~\cite{Kawaguchi:2008xi} 
supported
by $\pi_3$. 
In particular, superfluid vortices are one of the most important consequences of superfluidity;
when a superfluid is rotating, vortices 
aligned along the rotation axis are created. 
In a typical neutron star, the number of vortices 
can reach $10^{19}$. 
A singly quantized vortex 
carrying a unit circulation
has winding solely inside 
($2\pi$ rotation in) 
the baryon symmetry
$\U(1)_{\L+\R}$ 
in Eq.~\ref{eq:OPS-bs} which does not correspond to the minimal element of $\pi_1$. On the other hand, the minimally winding element of $\pi_1$ 
is given by a half-quantized Alice string 
carrying a half circulation, 
which is constructed by a 
$\pi$ rotation in $\U(1)_{\L+\R}$  
as well as a $\pi$ rotation in $\SO(3)_{\rm S}$~\cite{Leonhardt:2000km, Chatterjee:2017jsi}.
A singly quantized vortex can decay into a pair of half-quantized vortices depending on the model parameters.

\subsection{Dirac points} \label{sec:Dirac_points}

The anisotropy of the single-particle energy 
in Eqs.~\eqref{eq:energy_momentum_dispersion_Z2_negarive_negative_S} and \eqref{eq:energy_momentum_dispersion_Z2_negarive_negative_A} induces the existence of the Dirac fermions, i.e., the massless fermions with both left and right components.
The Dirac fermions exist at special points around the Fermi surface in the momentum space, as they are indicated by the cusps shown in the panels (a) and (c) in Fig.~\ref{fig:231230}.
Such points are called the Dirac points.
Here we focus on the Dirac fermions in the ${\cal A}1^{+}$ condensate directed along the $z$ axis: $\vect{\Delta}_{0}=\vect{\Delta}_{1}=\vect{\Delta}_{3}=\vect{0}$ and $\vect{\delta}=(0,0,\delta)$. A similar discussion can be applied to the ${\cal S}_{a}1^{-}$ condensates ($a=0,1,3$) as well.

From Eq.~\eqref{eq:general_Z2_negative_negative_generating_functional_connected}, we obtain the Hamiltonian given by
\begin{align}
   {\cal H}
&=
   \left(
   \begin{array}{cc}
      -\vect{p}\!\cdot\!\gamma^{0}\vect{\gamma} - \mu &
      - i \sum_{a} \gamma^{0}\vect{\gamma}\gamma^{5}\tau_{a} \vect{\Delta}_{a}
      - i \gamma^{0}\vect{\gamma}\tau_{2} \vect{\delta} \\ 
        i \sum_{a} \gamma^{0}\vect{\gamma}\gamma^{5}\tau_{a} \vect{\Delta}_{a}^{\dag}
      + i \gamma^{0}\vect{\gamma}\tau_{2} \vect{\delta}^{\dag} &
      - \vect{p}\!\cdot\!\gamma^{0}\vect{\gamma} + \mu
   \end{array}
   \right),
\label{eq:SU2PF_hamiltonian}
\end{align}
 in the Nambu-Gor'kov space
 for the given mean fields, $\vect{\Delta}_{a}$ and $\vect{\delta}$, for the ${\cal S}_{a}1^{-}$ and ${\cal A}1^{+}$ condensates, respectively.
 The eigenvalues of the Hamiltonian~\eqref{eq:SU2PF_hamiltonian} are given by
the single-particle energies given in Eqs.~\eqref{eq:energy_momentum_dispersion_Z2_negarive_negative_S} and \eqref{eq:energy_momentum_dispersion_Z2_negarive_negative_A}.
Considering the ${\cal A}1^{+}$ condensate, we find that the Hamiltonian~\eqref{eq:SU2PF_hamiltonian} is expressed by
\begin{align}
   {\cal H}_{\delta}(\vect{p})
&=
   \left(
   \begin{array}{cc}
       - \vect{p}\!\cdot\!\gamma^{0}\vect{\gamma} - \mu & -i\gamma^{0}\vect{\gamma}\tau_{2} \vect{\delta} \\ 
         i\gamma^{0}\vect{\gamma}\tau_{2} \vect{\delta}^{\dag} & -\vect{p}\!\cdot\!\gamma^{0}\vect{\gamma} + \mu
   \end{array}
   \right),
\label{eq:SU2PF_hamiltonian_ansatz}
\end{align}
and that
the Dirac points in ${\cal H}_{\delta}(\vect{p})$ are given by
\begin{align}
   \vect{p}^{\ast}_{\pm} = \bigl(0,0,\pm\sqrt{\mu^{2}+|\delta|^{2}}\bigr),
\label{eq:SU2PF_Dirac_point}
\end{align}
by the condition of the massless (zero energy) fermion, i.e., eigenvalues zero.
The Dirac points appear near the north and south poles on the Fermi surface.
Considering the eigenvalue problem for the Hamiltonian ${\cal H}_{\delta}(\vect{p}^{\ast}_{\pm})$ at the Dirac points, 
we obtain the eigenfunctions $U_{m}^{\pm}$ ($m=1,2,\dots,8$) for the Dirac fermions with chirality $\pm1$.
Because the Hamiltonian~\eqref{eq:SU2PF_hamiltonian_ansatz} is the $16\times16$ dimensional matrix in the Nambu-Gor'kov formalism,
$U^{\pm}_{m}$ contains the following components: 
\begin{align}
   U^{\pm}_{m}
=
   \left(
   \begin{array}{c}
      \psi_{1} \,\, \mathrm{particle} \,\, \mathrm{component} \\ 
      \psi_{2} \,\, \mathrm{particle} \,\, \mathrm{component} \\ 
      \psi_{1} \,\, \mathrm{antiparticle} \,\, \mathrm{component} \\ 
      \psi_{2} \,\, \mathrm{antiparticle} \,\, \mathrm{component}
   \end{array}
   \right).
\end{align}
The explicit forms of $U_{m}^{\pm}$ are shown in Eq.~\eqref{eq:SU2PF_hamiltonian_ansatz_Dirac_point_wf} in Appendix~\ref{sec:H3pm_DP_eigenfunctions}.

We investigate the fermionic excitation modes at low energy around the Dirac points ($\vect{p}^{\ast}_{\pm}$).
Using the eigenfunctions $U_{m}^{\pm}$ ($m=1,2,\dots,8$) in Eq.~\eqref{eq:SU2PF_hamiltonian_ansatz_Dirac_point_wf},
we define the effective Hamiltonian as an $8 \times 8$ dimensional matrix given by
\begin{align}
   \widetilde{\cal H}_{\delta}^{\ast\pm}(\vect{q})
=
   \left(
   \begin{array}{c}
      U_{m}^{\pm\dag} {\cal H}_{\delta}(\vect{p}) U^{\pm}_{n}
   \end{array}
   \right)_{m,n=1,\dots,8},
\label{eq:SU2PF_hamiltonian_ansatz_Dirac_point_eff_ham_def}
\end{align}
with a small deviation of the momentum from the Dirac points, $\vect{q}=\vect{p}-\vect{p}^{\ast}_{\pm}$.
Taking the lowest order for the expansion by $\vect{q}$, we find that the Hamiltonian~\eqref{eq:SU2PF_hamiltonian_ansatz_Dirac_point_eff_ham_def} is expressed by
\begin{align}
   \widetilde{\cal H}_{\delta}^{\ast\pm}(\vect{q})
=
   \left(
   \begin{array}{cccc}
      \tilde{\vect{q}}\!\cdot\!\vect{\sigma} & 0 & 0 & 0 \\
      0 & - \tilde{\vect{q}}\!\cdot\!\vect{\sigma} & 0 & 0 \\
      0 & 0 & \tilde{\vect{q}}\!\cdot\!\vect{\sigma} & 0 \\
      0 & 0 & 0 & - \tilde{\vect{q}}\!\cdot\!\vect{\sigma}
   \end{array}
   \right).
\label{eq:SU2PF_hamiltonian_ansatz_Dirac_point_eff_ham}
\end{align}
Here, instead of $\vect{q}$, we define $ \tilde{\vect{q}}$ by
\begin{align}
   \tilde{\vect{q}}
= \bigl(\tilde{q}_{1},\tilde{q}_{2},\tilde{q}_{3}\bigr)
=
   \left(
    - \dfrac{q_{1}}{\sqrt{1+\dfrac{\mu^{2}}{|\delta|^{2}}}}, - \dfrac{q_{2}}{\sqrt{1+\dfrac{\mu^{2}}{|\delta|^{2}}}}, q_{3}
   \right),
\label{eq:tilde_q_def}
\end{align}
for the convenience of the notation.
The momenta in the direction along the $x$ and $y$ axes, $\tilde{q}_{1}$ and $\tilde{q}_{2}$, are scaled by the factor $1/\sqrt{1+\mu^{2}/|\delta|^{2}}$ against the original momenta, $q_{1}$ and $q_{2}$.
There is no modification for the $z$ axis.
The submatrices, $\tilde{\vect{q}}\cdot\vect{\sigma}$ and $-\tilde{\vect{q}}\cdot\vect{\sigma}$, in the Hamiltonian~\eqref{eq:SU2PF_hamiltonian_ansatz_Dirac_point_eff_ham} indicate the fermionic excitation modes with negative and positive parities, respectively, near the Dirac points.
The eigenvalues of the Hamiltonian~\eqref{eq:SU2PF_hamiltonian_ansatz_Dirac_point_eff_ham} are given by
\begin{align}
 \pm \varepsilon_{\tilde{\vect{q}}}
= \pm |\tilde{\vect{q}}|
= \pm \sqrt{ \dfrac{q_{1}^{2}+q_{2}^{2}}{1+\dfrac{\mu^{2}}{|\delta|^{2}}} + q_{3}^{2}},
\label{eq:SU2PF_hamiltonian_ansatz_Dirac_point_eff_ham_sol}
\end{align}
with four degenerate solutions in each $\pm$.
The dispersion relation~\eqref{eq:SU2PF_hamiltonian_ansatz_Dirac_point_eff_ham_sol} shows a special property of the fermionic excitation modes at the Dirac points:
the propagation speed of the fermionic excitations on the $x$-$y$ planes is slower than the speed of light, 
while the propagation speed in the direction of the $z$ axis is the same as the speed of light.
The former is because the slope in the energy-momentum dispersion relation is very small ($|\delta| \ll \mu$) in the $x$ and $y$ directions.
Notice that the vector condensate is directed along the $z$ axis in the present setting.
Therefore, the fermionic excitation modes have a tendency to propagate on the $x$-$y$ plane rather than the $z$ axis, because the energy cost of excitation is smaller on the $x$-$y$ plane.
This result shows the anisotropy of the transport phenomenon in the emergent chiral superfluids.

\section{Conclusion and outlook} \label{sec:conclusion}

We have discussed the emergent chiral superfluids of neutron matter, where the neutron $n(940)$ (mass 940 MeV, spin-parity
$J^{P}=1/2^{+}$) and the chiral partner $n^{\ast}(1535)$ (mass 1535 MeV, spin-parity $J^{P}=1/2^{-}$) are degenerate forming parity doubling at sufficiently high density.
Apart from the usual chiral symmetry, we have introduced the $G_{\L} \times G_{\R} = \U(1)_{(1-2)\L} \times \U(1)_{(1-2)\R}$
and  $\SU(2)_{\L} \times \SU(2)_{\R}$ emergent chiral symmetries as proper symmetries in the parity-doublet model since the naive and mirror assignments are naturally included as their subgroups.
The emergent chiral symmetries are novel symmetries proposed in this study for the first time.
Introducing the four-point interactions for $n$ and $n^{\ast}$, we have classified the Lagrangian in each emergent chiral symmetry and analyzed the Cooper pairings based on the Nambu-Gor'kov formalism.
The emergent chiral superfluids are characterized by several different condensates: the emergent-chiral-symmetric and vector condensate (${\cal S}_{a}1^{-}$; $a=0,1,3$) and the emergent-chiral-asymmetric and axialvector condensate (${\cal A}1^{+}$).
For the $\U(1)_{(1-2)\L} \times \U(1)_{(1-2)\R}$ emergent chiral symmetry, we have obtained (i) either ${\cal S}_{0}1^{-}$ or ${\cal S}_{3}1^{-}$ and (ii) either ${\cal S}_{1}1^{-}$ or ${\cal A}1^{+}$ according to the relative strengths in the interaction terms.
For the $\SU(2)_{\L} \times \SU(2)_{\R}$ emergent chiral symmetry, one of the ${\cal S}_{0}1^{-}$, ${\cal S}_{1}1^{-}$, ${\cal S}_{3}1^{-}$ and ${\cal A}1^{+}$ condensates is formed.
The emergent chiral symmetry as well as the spatial rotational symmetry are broken by these condensates.
We have analyzed the patterns of symmetry breaking and the appearance of the NG bosons consisting of six quarks, and found that the emergent chiral superfluids exhibit several topological excitations according to the homotopy groups.
We have also shown the existence of the (massless) Dirac fermions at Dirac points in momentum space, indicating an anisotropy in the transport phenomena.

In this paper, we have considered 
condensates of only one of 
the vector condensations 
$\vect{\Delta}_{a}$ ($a=0,1,3$) and $\vect{\delta}$ for 
${\cal S}_{0}1^{-}$, 
${\cal S}_{1}1^{-}$,
${\cal S}_{3}1^{-}$ and ${\cal A}1^{+}$ for simplicity. 
We have further assumed for each case 
a condensation of only 
one spatial component of each three-dimensional vector 
condensation, 
$\vect{\Delta}_{a}=(0,0,\Delta_{a})$ or $\vect{\delta}=(0,0,\delta)$,
which is not the most general.
This situation
corresponds to the polar phase 
in spinor BECs, 
in which case depending on couplings
there is also another phase, the so-called ferromagnetic phase 
where the $\SO(3)$ rotational symmetry is completely 
broken spontaneously 
with magnetizations~\cite{Kawaguchi:2012ii}. 
Investigating a possibility of such a  
ferromagnetic phase 
with a possible origin of magnetic fields in neutron stars  
is one of future directions.
It is also interesting to study whether 
a simultaneous condensation of
${\cal S}_{0}1^{-}$ and 
${\cal S}_{3}1^{-}$ 
and that of 
${\cal S}_{1}1^{-}$ and ${\cal A}1^{+}$ 
can occur in general. 
In the ${\cal S}_{0}1^{-}$ or 
${\cal S}_{3}1^{-}$ phase 
in Fig.~\ref{fig:240902_phase_diagram}, 
the ground states  of 
the condensates 
${\cal S}_{0}1^{-}$ 
and 
${\cal S}_{3}1^{-}$ 
are energetically degenerated, 
while 
in the 
${\cal S}_{1}1^{-}$ or ${\cal A}1^{+}$ phase
those of 
${\cal S}_{1}1^{-}$ and ${\cal A}1^{+}$ 
are degenerated.
Thus, there may exist domain walls 
interpolating these two condensates in each phase.

As discussed in Sec.~\ref{sec:DSB}, 
there exists a rich variety of topological objects in dynamical symmetry breakings in our superfluids,
among which a particularly important class is given by superfluid vortices; 
since neutron stars are rapidly rotating, 
there appear a large number of 
superfluid vortices along the rotation axis 
and they may form an Abrikosov's lattice. 
In our case, a singly quantized vortex may 
be split into a pair of 
half-quantized Alice strings~\cite{Leonhardt:2000km, Chatterjee:2017jsi}.
Such superfluid vortices may be important for pulsar glitches, 
as was proposed for $^3P_2$-$^1S_0$ interface~\cite{Marmorini:2020zfp}. 
If our  superfluids are connected to 
$^3 P_2$ neutron superfluids 
as illustrated in Fig.~\ref{fig:240807_star_inside}, 
half-quantized Alice strings in the former 
may be connected to 
half-quantized non-Abelian vortices~\cite{Masuda:2016vak,Kobayashi:2022moc,Kobayashi:2022dae,Masaki:2021hmk}
in the latter.

Anisotropic transport of Dirac fermions may have
an impact on neutron star cooling. It is an interesting subject to study the interaction between neutrinos and Dirac fermions. There we will also need to consider the NG bosons and study the coupled system among neutrinos, Dirac fermions and NG bosons.
The Berry phase around the Dirac cone will lead to a nontrivial topological charge showing properties of topological materials.

Studying the NG bosons consisting of six quarks in the 
emergent chiral superfluids
may be an interesting subject for multiquark systems in the light of exotic hadrons
by the quark number counting.
Similar situations can be considered in the neutron $^{1}S_{0}$ and $^{3}P_{2}$ superfluids.
Revealing the properties of these NG bosons, such as dispersion relations, interactions with medium and so on, at high density may give us a hint to understand multiquark systems in a way different from the conventional hadron spectroscopy in vacuum.
Investigating contributions of these NG bosons to neutron star cooling is also 
an important future problem.

Our results in this paper are obtained exclusively with the specific part of the entire Lagrangian, $\mathcal{L}_{\rm com}$. The given classifications of symmetries and operators both in the naive and mirror assignments may suggest further richness of the phase structure in a more general situation. 
It is of particular interest to explore the full theory to possibly pin down any characteristics as distinct properties of the given chirality assignment in high-density QCD.
Besides, it may also be intriguing to clarify a potential link of the emergent chiral symmetry to the chiral spin symmetry proposed for the chiral structure of hadrons in hot and dense matter~\cite{Glozman:2014mka,Glozman:2022lda,Glozman:2022zpy,Glozman:2015qva}, as well as to the quarkyonic phase allowing the hadronic degrees freedom near the Fermi surface of quark matter~\cite{McLerran:2007qj}.
Those open problems are left for the future studies.

\section*{Acknowledgments}
This work is supported
by the World Premier International Research Center Initiative (WPI) in the Ministry of Education, Culture, Sports, Science (MEXT), Japan.
The work of M.N. is supported in part by 
 JSPS KAKENHI [Grants  No. JP22H01221 (M.N.)]. 
C.S. acknowledges the support of the Polish National Science Centre (NCN) under OPUS Grant No.~2022/45/B/ST2/01527.

\appendix

\section{Naive and mirror assignments and extension of chiral symmetry} \label{sec:extension_chiral_symmetry}

We consider two baryon fields denoted by $\psi_{1}$ and $\psi_{2}$ as chiral doublet.
$\psi_{i}$ ($i=1,2$) has the left and right components according to chirality, $\psi_{i\L}$ and $\psi_{i\R}$.
Thus, $\psi_{i}$ ($i=1,2$) is expressed by a direct sum of $\psi_{i\L}$ and $\psi_{i\R}$:
\begin{align}
   \psi_{1} = \psi_{1\L} + \psi_{1\R}, \quad 
   \psi_{2} = \psi_{2\L} + \psi_{2\R}.
\end{align}
The left and right components, $\psi_{i\R}$ and $\psi_{i\L}$ ($i=1,2$), are expressed by the projection operators
\begin{align}
   \psi_{1\L} = \frac{1-\gamma_{5}}{2} \psi_{1}, \quad 
   \psi_{2\L} = \frac{1-\gamma_{5}}{2} \psi_{2}, \quad 
   \psi_{1\R} = \frac{1+\gamma_{5}}{2} \psi_{1}, \quad 
   \psi_{2\R} = \frac{1+\gamma_{5}}{2} \psi_{2}.
\end{align}
In $\psi_{1}$ and $\psi_{2}$, the left and right components are transformed under the chiral symmetry $\U(1)_{\L} \times \U(1)_{\R}$.
The linear combinations of $\psi_{1}$ and $\psi_{2}$ make the basis for the fields of the nucleon ($N$) with spin-parity $J^{P}=1/2^{+}$ and the excited nucleon ($N^{\ast}(1535)$) with $J^{P}=1/2^{-}$.

The chiral symmetry $\U(1)_{\L} \times \U(1)_{\R}$ indicates usually the simultaneous transformations as a global symmetry:
\begin{align}
   \psi_{1\L} \rightarrow U_{\L}\psi_{1\L}, \quad 
   \psi_{2\L} \rightarrow U_{\L}\psi_{2\L}, \quad 
   \psi_{1\R} \rightarrow U_{\R}\psi_{1\R}, \quad 
   \psi_{2\R} \rightarrow U_{\R}\psi_{2\R},
\label{eq:chiral_transformation_naive}
\end{align}
for $U_{\L} \in \U(1)_{\L}$ and $U_{\R} \in \U(1)_{\R}$.
In this case, the transformation rules are common to $\psi_{i\L}$ and $\psi_{i\R}$ ($i=1,2$), respectively, for each chirality:
the left component transforms under $\U(1)_{\L}$ and the right component transforms under $\U(1)_{\R}$.
This is, however, not the unique transformation.
Another possible transformation in chiral symmetry would be given by
\begin{align}
   \psi_{1\L} \rightarrow U_{\L}\psi_{1\L}, \quad 
   \psi_{2\L} \rightarrow U_{\R}\psi_{2\L}, \quad 
   \psi_{1\R} \rightarrow U_{\R}\psi_{1\R}, \quad 
   \psi_{2\R} \rightarrow U_{\L}\psi_{2\R},
\label{eq:chiral_transformation_mirror}
\end{align}
where $\psi_{1\L/\R}$ transforms under $\U(1)_{\L/\R}$ and $\psi_{2\L/\R}$ transforms under $\U(1)_{\R/\L}$.
Notice in this case that the chiral transformation of $\psi_{1}$ is different from that of $\psi_{2}$.
The chiral transformations~\eqref{eq:chiral_transformation_naive} and \eqref{eq:chiral_transformation_mirror} are called the {\it naive} and {\it mirror} assignments, respectively, in the chiral doublet formalism.
It is a nontrivial problem which of naive and mirror assignments are realized in the QCD, and this problem is researched by several researchers.

Including the naive and mirror assignments in the chiral symmetry, as discussed in Sec.~\ref{sec:emergent chiral_symmetry},
we consider the emergent chiral symmetry,  $\U(1)_{1\L}\times\U(1)_{2\L}\times\U(1)_{1\R}\times\U(1)_{2\R}$, where $\U(1)_{i\alpha}$ ($i=1,2$ and $\alpha=\L,\R$) indicates the $\U(1)$ symmetry for each $\psi_{i\alpha}$.
In this symmetry, $\psi_{i\alpha}$ is transformed in a completely independent manner by
\begin{align}
   \psi_{1\L} \rightarrow U_{1\L} \psi_{1\L}, \quad 
   \psi_{2\L} \rightarrow U_{2\L} \psi_{2\L}, \quad 
   \psi_{1\R} \rightarrow U_{1\R} \psi_{1\R}, \quad 
   \psi_{2\R} \rightarrow U_{2\R} \psi_{2\R},
\end{align}
with  $U_{1\L} \in \U(1)_{1\L}$, $U_{2\L} \in \U(1)_{2\L}$, $U_{1\R} \in \U(1)_{1\R}$ and $U_{2\R} \in \U(1)_{2\R}$.
The naive symmetry is reproduced by setting $U_{1\L}=U_{2\L}$ and $U_{1\R}=U_{2\R}$.
Similarly, the mirror symmetry is reproduced by setting $U_{1\L}=U_{2\R}$ and $U_{1\R}=U_{2\L}$.

\section{Lagrangians in naive and mirror assignments} \label{sec:Lagrangians_naive_mirror}

In Sec.~\ref{sec:formalism}, we have introduced the Lagrangian with four-point interaction ${\cal L}_{\mathrm{com}}$, which is commonly included both in the naive and mirror assignments.
In this Appendix, we show the Lagrangian belonging to each of the naive assignments and the mirror assignments.
The Lagrangian in the naive assignment is given by
\begin{align}
   {\cal L}_{\naive}
&= 
   \bar{\psi}_{1} i \gamma\partial \psi_{1} + \bar{\psi}_{2} i \gamma\partial \psi_{2}
+ g_{1111} \frac{1}{4}
\bigl( (\bar{\psi}_{1}\psi_{1})^{2} + (\bar{\psi}_{1}i\gamma_{5}\psi_{1})^{2} \bigr)
+ g_{2222} \frac{1}{4} \bigl( (\bar{\psi}_{2}\psi_{2})^{2} + (\bar{\psi}_{2}i\gamma_{5}\psi_{2})^{2} \bigr)
\nonumber \\ & 
+ g_{1221} \frac{1}{2}
   \bigl( (\bar{\psi}_{1}\psi_{2})(\bar{\psi}_{2}\psi_{1}) + (\bar{\psi}_{1}i\gamma_{5}\psi_{2})(\bar{\psi}_{2}i\gamma_{5}\psi_{1}) \bigr)
   \nonumber \\ & 
+ g_{1122} \frac{1}{2}
   \bigl( (\bar{\psi}_{1}\psi_{1})(\bar{\psi}_{2}\psi_{2}) + (\bar{\psi}_{1}i\gamma_{5}\psi_{1})(\bar{\psi}_{2}i\gamma_{5}\psi_{2}) \bigr)
   \nonumber \\ & 
+ g_{1112} \frac{1}{2}
   \bigl(
         (\bar{\psi}_{1}\psi_{1}) (\bar{\psi}_{1}\psi_{2})
      + (\bar{\psi}_{1}\psi_{1}) (\bar{\psi}_{2}\psi_{1})
       + (\bar{\psi}_{1}i\gamma_{5}\psi_{1}) (\bar{\psi}_{1}i\gamma_{5}\psi_{2})
       + (\bar{\psi}_{1}i\gamma_{5}\psi_{1}) (\bar{\psi}_{2}i\gamma_{5}\psi_{1})
   \bigr)
   \nonumber \\ & 
+ g_{1222} \frac{1}{2}
   \bigl(
         (\bar{\psi}_{1}\psi_{2}) (\bar{\psi}_{2}\psi_{2})
      + (\bar{\psi}_{2}\psi_{1}) (\bar{\psi}_{2}\psi_{2})
       + (\bar{\psi}_{1}i\gamma_{5}\psi_{2}) (\bar{\psi}_{2}i\gamma_{5}\psi_{2})
      + (\bar{\psi}_{2}i\gamma_{5}\psi_{1}) (\bar{\psi}_{2}i\gamma_{5}\psi_{2})
   \bigr)
   \nonumber \\ & 
+ g_{1212} \frac{1}{4}
   \bigl(
         (\bar{\psi}_{1}\psi_{2}) (\bar{\psi}_{1}\psi_{2})
      + (\bar{\psi}_{2}\psi_{1}) (\bar{\psi}_{2}\psi_{1})
       + (\bar{\psi}_{1}i\gamma_{5}\psi_{2}) (\bar{\psi}_{1}i\gamma_{5}\psi_{2})
       + (\bar{\psi}_{2}i\gamma_{5}\psi_{1}) (\bar{\psi}_{2}i\gamma_{5}\psi_{1})
   \bigr).
\label{eq:Lagrangian_naive_summary}
\end{align}
The Lagrangian in the mirror assignment is given by
\begin{align}
   {\cal L}_{\mirror}
&=
   \bar{\psi}_{1} i \gamma\partial \psi_{1} + \bar{\psi}_{2} i \gamma\partial \psi_{2}
+ g_{1111} \frac{1}{4} \bigl( (\bar{\psi}_{1}\psi_{1})^{2} + (\bar{\psi}_{1}i\gamma_{5}\psi_{1})^{2} \bigr)
+ g_{2222} \frac{1}{4} \bigl( (\bar{\psi}_{2}\psi_{2})^{2} + (\bar{\psi}_{2}i\gamma_{5}\psi_{2})^{2} \bigr)
\nonumber \\ & 
+ g_{1221} \frac{1}{2}
   \bigl( (\bar{\psi}_{1}\psi_{2})(\bar{\psi}_{2}\psi_{1}) + (\bar{\psi}_{1}i\gamma_{5}\psi_{2})(\bar{\psi}_{2}i\gamma_{5}\psi_{1}) \bigr)
   \nonumber \\ & 
+ f_{1122} \frac{1}{2}
   \bigl( (\bar{\psi}_{1}\psi_{1})(\bar{\psi}_{2}\psi_{2}) - (\bar{\psi}_{1}i\gamma_{5}\psi_{1})(\bar{\psi}_{2}i\gamma_{5}\psi_{2}) \bigr)
   \nonumber \\ & 
+ f_{1212} \frac{1}{2}
   \bigl(
         (\bar{\psi}_{1}\psi_{2}) (\bar{\psi}_{1}\psi_{2})
      + (\bar{\psi}_{2}\psi_{1}) (\bar{\psi}_{2}\psi_{1})
      - (\bar{\psi}_{1}i\gamma_{5}\psi_{2}) (\bar{\psi}_{1}i\gamma_{5}\psi_{2})
      - (\bar{\psi}_{2}i\gamma_{5}\psi_{1}) (\bar{\psi}_{2}i\gamma_{5}\psi_{1})
   \bigr)
   \nonumber \\ & 
+ f_{1221} \frac{1}{2}
   \bigl( (\bar{\psi}_{1}\psi_{2})(\bar{\psi}_{2}\psi_{1}) - (\bar{\psi}_{1}i\gamma_{5}\psi_{2})(\bar{\psi}_{2}i\gamma_{5}\psi_{1}) \bigr)
   \nonumber \\ & 
+ g_{1212} \frac{1}{4}
   \bigl(
         (\bar{\psi}_{1}\psi_{2}) (\bar{\psi}_{1}\psi_{2})
      + (\bar{\psi}_{2}\psi_{1}) (\bar{\psi}_{2}\psi_{1})
       + (\bar{\psi}_{1}i\gamma_{5}\psi_{2}) (\bar{\psi}_{1}i\gamma_{5}\psi_{2})
       + (\bar{\psi}_{2}i\gamma_{5}\psi_{1}) (\bar{\psi}_{2}i\gamma_{5}\psi_{1})
   \bigr)
   \nonumber \\ & 
 - m_{0} \bigl( \bar{\psi}_{1}\psi_{2} + \bar{\psi}_{2}\psi_{1} \bigr).
\label{eq:Lagrangian_mirror_summary}
\end{align}
In Eqs.~\eqref{eq:Lagrangian_naive_summary} and \eqref{eq:Lagrangian_mirror_summary}, we need to set $g_{1111}=-16g_{\perp}$, $g_{2222}=-16g_{\perp}'$ and $g_{1221}=-16g_{\parallel}$ to be consistent with the coupling constants in  Eq.~\eqref{eq:Lagrangian_general_summary}.
The coupling constants in Eqs.~\eqref{eq:Lagrangian_naive_summary} and \eqref{eq:Lagrangian_mirror_summary} are regarded as free parameters.
In the last term in Eq.~\eqref{eq:Lagrangian_mirror_summary}, $m_{0}$ indicates the mass parameter which does not break the chiral symmetry.

\section{Fierz identities} \label{sec:Fierz_transformation}

We summarize the Fierz identities for the Dirac matrices and the generators of $\SU(N)$ symmetry (see, e.g., Ref.~\cite{Buballa:2003qv}).
They are used to calculate the Fierz transformation by rearranging the interaction term $(\bar{\psi}\Gamma_{\mu a}\psi)(\bar{\psi}\Gamma_{\mu a}\psi)$ for a matrix $\Gamma_{\mu a}$ with $\mu=0,1,2,3$ for the Lorentz indices and $a=1,\dots,N^{2}-1$ for the $\SU(N)$  symmetry.

\subsection{Dirac matrices} \label{sec:Feirz_Dirac}

The Fierz identities for the particle-antiparticle channels are given by
\begin{align}
   (\mathbbm{1})_{ij} (\mathbbm{1})_{kl}
&=
   \frac{1}{4} (\mathbbm{1})_{il} (\mathbbm{1})_{kj}
 - \frac{1}{4} (i\gamma_{5})_{il} (i\gamma_{5})_{kj}
+ \frac{1}{4} (\gamma^{\mu})_{il} (\gamma_{\mu})_{kj}
 - \frac{1}{4} (\gamma^{\mu}\gamma_{5})_{il} (\gamma_{\mu}\gamma_{5})_{kj}
+ \frac{1}{8} (\sigma^{\mu\nu})_{il} (\sigma_{\mu\nu})_{kj},
\nonumber \\ 
   (i\gamma_{5})_{ij} (i\gamma_{5})_{kl}
&=
 - \frac{1}{4} (\mathbbm{1})_{il} (\mathbbm{1})_{kj}
+ \frac{1}{4} (i\gamma_{5})_{il} (i\gamma_{5})_{kj}
+ \frac{1}{4} (\gamma^{\mu})_{il} (\gamma_{\mu})_{kj}
 - \frac{1}{4} (\gamma^{\mu}\gamma_{5})_{il} (\gamma_{\mu}\gamma_{5})_{kj}
 - \frac{1}{8} (\sigma^{\mu\nu})_{il} (\sigma_{\mu\nu})_{kj},
\nonumber \\ 
   (\gamma^{\mu})_{ij} (\gamma_{\mu})_{kl}
&=
   (\mathbbm{1})_{il} (\mathbbm{1})_{kj}
+ (i\gamma_{5})_{il} (i\gamma_{5})_{kj}
 - \frac{1}{2} (\gamma^{\mu})_{il} (\gamma_{\mu})_{kj}
 - \frac{1}{2} (\gamma^{\mu}\gamma_{5})_{il} (\gamma_{\mu}\gamma_{5})_{kj},
\nonumber \\ 
   (\gamma^{\mu}\gamma_{5})_{ij} (\gamma_{\mu}\gamma_{5})_{kl}
&=
 - (\mathbbm{1})_{il} (\mathbbm{1})_{kj}
 - (i\gamma_{5})_{il} (i\gamma_{5})_{kj}
 - \frac{1}{2} (\gamma^{\mu})_{il} (\gamma_{\mu})_{kj}
 - \frac{1}{2} (\gamma^{\mu}\gamma_{5})_{il} (\gamma_{\mu}\gamma_{5})_{kj},
\nonumber \\ 
   (\sigma^{\mu\nu})_{ij} (\sigma_{\mu\nu})_{kl}
&=
   3 (\mathbbm{1})_{il} (\mathbbm{1})_{kj}
 - 3 (i\gamma_{5})_{il} (i\gamma_{5})_{kj}
 - \frac{1}{2} (\sigma^{\mu\nu})_{il} (\sigma_{\mu\nu})_{kj},
\label{eq:Fierz_Dirac_pa_1}
\end{align}
and
\begin{align}
   (\gamma^{0})_{ij} (\gamma^{0})_{kl}
&=
   \frac{1}{4} (\mathbbm{1})_{il} (\mathbbm{1})_{kj}
+ \frac{1}{4} (i\gamma_{5})_{il} (i\gamma_{5})_{kj}
+ \frac{1}{4} (\gamma^{0})_{il} (\gamma^{0})_{kj}
 - \frac{1}{4} (\gamma^{m})_{il} (\gamma_{m})_{kj}
+ \frac{1}{4} (\gamma^{0}\gamma_{5})_{il} (\gamma^{0}\gamma_{5})_{kj}
 - \frac{1}{4} (\gamma^{m}\gamma_{5})_{il} (\gamma_{m}\gamma_{5})_{kj}
    \nonumber \\ & 
 - \frac{1}{4} (\sigma^{0n})_{il} (\sigma_{0n})_{kj}
 + \frac{1}{8} (\sigma^{mn})_{il} (\sigma_{mn})_{kj},
\label{eq:Fierz_Dirac_pa_2}
\end{align}
with $\mu,\nu=0,1,2,3$ and $i,j,k,l,m,n=1,2,3$.
$\mathbbm{1}$ is the $4\times4$ unit matrix.
The repeated indices are implicitly assumed to make sums.
The Fierz identities for the particle-particle channels are given by
\begin{align}
   (\mathbbm{1})_{ij} (\mathbbm{1})_{kl}
&=
   \frac{1}{4} (i\gamma_{5}C)_{ik} (Ci\gamma_{5})_{lj}
 - \frac{1}{4} (C)_{ik} (C)_{lj}
+ \frac{1}{4} (\gamma^{\mu}\gamma_{5}C)_{ik} (C\gamma_{\mu}\gamma_{5})_{lj}
 - \frac{1}{4} (\gamma^{\mu}C)_{ik} (C\gamma_{\mu})_{lj}
 - \frac{1}{8} (\sigma^{\mu\nu}C)_{ik} (C\sigma_{\mu\nu})_{lj},
\nonumber \\ 
   (i\gamma_{5})_{ij} (i\gamma_{5})_{kl}
&=
 - \frac{1}{4} (i\gamma_{5}C)_{ik} (Ci\gamma_{5})_{lj}
+ \frac{1}{4} (C)_{ik} (C)_{lj}
+ \frac{1}{4} (\gamma^{\mu}\gamma_{5}C)_{ik} (C\gamma_{\mu}\gamma_{5})_{lj}
 - \frac{1}{4} (\gamma^{\mu}C)_{ik} (C\gamma_{\mu})_{lj}
+ \frac{1}{8} (\sigma^{\mu\nu}C)_{ik} (C\sigma_{\mu\nu})_{lj},
\nonumber \\ 
   (\gamma^{\mu})_{ij} (\gamma_{\mu})_{kl}
&=
   (i\gamma_{5}C)_{ik} (Ci\gamma_{5})_{lj}
+ (C)_{ik} (C)_{lj}
 - \frac{1}{2} (\gamma^{\mu}\gamma_{5}C)_{ik} (C\gamma_{\mu}\gamma_{5})_{lj}
 - \frac{1}{2} (\gamma^{\mu}C)_{ik} (C\gamma_{\mu})_{lj},
\nonumber \\ 
   (\gamma^{\mu}\gamma_{5})_{ij} (\gamma_{\mu}\gamma_{5})_{kl}
&=
   (i\gamma_{5}C)_{ik} (Ci\gamma_{5})_{lj}
+ (C)_{ik} (C)_{lj}
+ \frac{1}{2} (\gamma^{\mu}\gamma_{5}C)_{ik} (C\gamma_{\mu}\gamma_{5})_{lj}
+ \frac{1}{2} (\gamma^{\mu}C)_{ik} (C\gamma_{\mu})_{lj},
\nonumber \\ 
   (\sigma^{\mu\nu})_{ij} (\sigma_{\mu\nu})_{kl}
&=
 - 3 (i\gamma_{5}C)_{ik} (Ci\gamma_{5})_{lj}
+ 3 (C)_{ik} (C)_{lj}
 - \frac{1}{2} (\sigma^{\mu\nu}C)_{ik} (C\sigma_{\mu\nu})_{lj},
\label{eq:Fierz_Dirac_pp_1}
\end{align}
and
\begin{align}
   (\gamma^{0})_{ij} (\gamma^{0})_{kl}
&=
   \frac{1}{4} (i\gamma_{5}C)_{il} (Ci\gamma_{5})_{kj}
+ \frac{1}{4} (C)_{il} (C)_{kj}
+ \frac{1}{4} (\gamma^{0}\gamma_{5}C)_{il} (C\gamma^{0}\gamma_{5})_{kj}
 - \frac{1}{4} (\gamma^{m}\gamma_{5}C)_{il} (C\gamma_{m}\gamma_{5})_{kj}
    \nonumber \\ & 
+ \frac{1}{4} (\gamma^{0}C)_{il} (C\gamma^{0})_{kj}
 - \frac{1}{4} (\gamma^{m}C)_{il} (C\gamma_{m})_{kj}
 - \frac{1}{4} (\sigma^{0n}C)_{il} (C\sigma_{0n})_{kj}
 + \frac{1}{8} (\sigma^{mn}C)_{il} (C\sigma_{mn})_{kj},
\label{eq:Fierz_Dirac_pp_2}
\end{align}
with $C=i\gamma^{2}\gamma^{0}$.

\subsection{$\SU(N)$ symmetry} \label{sec:Fierz_SUN}

We summarise the Fierz identities from the $\SU(N)$ generators.
The generators of $\SU(N)$ group are given by $\tau_{a}$ ($a=1,2,\dots,N^{2}-1$), which are normalized as $\tr(\tau_{a}\tau_{b})=2\delta_{ab}$.
$\mathbbm{1}$ is the unit matrix with $N \times N$ dimension, and $\tau_{0}=\sqrt{2/N}\mathbbm{1}$.
The Fierz identities for the particle-antiparticle channels are given by
\begin{align}
   (\mathbbm{1})_{ij} (\mathbbm{1})_{kl}
&=
   \frac{1}{N} (\mathbbm{1})_{il} (\mathbbm{1})_{kj}
+ \frac{1}{2} (\tau_{a})_{il} (\tau_{a})_{kj},
\nonumber \\ 
   (\tau_{a})_{ij} (\tau_{a})_{kl}
&=
   2\biggl(1-\frac{1}{N^{2}}\biggr) (\mathbbm{1})_{il} (\mathbbm{1})_{kj}
 - \frac{1}{N} (\tau_{a})_{il} (\tau_{a})_{kj},
\label{eq:Fierz_SUN_pa}
\end{align}
with $i,j,k,l=1,2,\dots,N$.
The Fierz identities for the particle-particle channels are given by
\begin{align}
   (\mathbbm{1})_{ij} (\mathbbm{1})_{kl}
&=
   \frac{1}{2} (\tau_{S})_{ik} (\tau_{S})_{lj}
+ \frac{1}{2} (\tau_{A})_{ik} (\tau_{A})_{lj},
\nonumber \\ 
   (\tau_{a})_{ij} (\tau_{a})_{kl}
&=
   \biggl(1-\frac{1}{N}\biggr) (\tau_{S})_{ik} (\tau_{S})_{lj}
 - \biggl(1+\frac{1}{N}\biggr) (\tau_{A})_{ik} (\tau_{A})_{lj}.
\label{eq:Fierz_SUN_pp}
\end{align}
The repeated indices are implicitly assumed to make sums.
$\tau_{S}$ are the symmetric matrices, and $\tau_{A}$ are the antisymmetric matrices.

\section{Eigenfunctions of Eq.~\eqref{eq:SU2PF_hamiltonian_ansatz}} \label{sec:H3pm_DP_eigenfunctions}

The eigenfunctions of ${\cal H}_{\delta}(\vect{p}^{\ast})$ in Eq.~\eqref{eq:SU2PF_hamiltonian_ansatz} are given by
\begin{align}
   U^{\pm}_{1}
&=
   \left(
   \begin{array}{c}
      0 \\
      0 \\
      0 \\
      \dfrac{i\delta\bigl(\mu\pm\sqrt{|\delta|^{2}+\mu^{2}}\bigr)}
      {|\delta|\sqrt{|\delta|^{2}+\bigl(\mu\pm\sqrt{|\delta|^{2}+\mu^{2}}\bigr)^{2}}} \\ 
      \hline
      0 \\
      0 \\
      0 \\
      0 \\ 
      \hline
      0 \\
      0 \\
      0 \\
      0 \\ 
      \hline
      0 \\
      0 \\
      0 \\
      \dfrac{|\delta|^{2}}{|\delta|\sqrt{|\delta|^{2}+\bigl(\mu\pm\sqrt{|\delta|^{2}+\mu^{2}}\bigr)^{2}}}
   \end{array}
   \right),
\quad 
   U^{\pm}_{2}
=
   \left(
   \begin{array}{c}
      0 \\
      0 \\
      \dfrac{i\delta\bigl(-\mu\pm\sqrt{|\delta|^{2}+\mu^{2}}\bigr)}
      {|\delta|\sqrt{|\delta|^{2}+\bigl(-\mu\pm\sqrt{|\delta|^{2}+\mu^{2}}\bigr)^{2}}} \\
      0 \\ 
      \hline
      0 \\
      0 \\
      0 \\
      0 \\ 
      \hline
      0 \\
      0 \\
      0 \\
      0 \\ 
      \hline
      0 \\
      0 \\
      \dfrac{|\delta|^{2}}{|\delta|\sqrt{|\delta|^{2}+\bigl(-\mu\pm\sqrt{|\delta|^{2}+\mu^{2}}\bigr)^{2}}} \\
      0
   \end{array}
   \right),
\nonumber \\ 
   U^{\pm}_{3}
&=
   \left(
   \begin{array}{c}
      0 \\
      \dfrac{i\delta\bigl(-\mu\pm\sqrt{|\delta|^{2}+\mu^{2}}\bigr)}
      {|\delta|\sqrt{|\delta|^{2}+\bigl(-\mu\pm\sqrt{|\delta|^{2}+\mu^{2}}\bigr)^{2}}} \\
      0 \\
      0 \\ 
      \hline
      0 \\
      0 \\
      0 \\
      0 \\ 
      \hline
      0 \\
      0 \\
      0 \\
      0 \\ 
      \hline
      0 \\
      \dfrac{|\delta|^{2}}{|\delta|\sqrt{|\delta|^{2}+\bigl(-\mu\pm\sqrt{|\delta|^{2}+\mu^{2}}\bigr)^{2}}} \\
      0 \\
      0
   \end{array}
   \right),
\quad 
   U^{\pm}_{4}
=
   \left(
   \begin{array}{c}
      \dfrac{i\delta\bigl(\mu\pm\sqrt{|\delta|^{2}+\mu^{2}}\bigr)}
      {|\delta|\sqrt{|\delta|^{2}+\bigl(\mu\pm\sqrt{|\delta|^{2}+\mu^{2}}\bigr)^{2}}} \\
      0 \\
      0 \\
      0 \\ 
      \hline
      0 \\
      0 \\
      0 \\
      0 \\ 
      \hline
      0 \\
      0 \\
      0 \\
      0 \\ 
      \hline
      \dfrac{|\delta|^{2}}{|\delta|\sqrt{|\delta|^{2}+\bigl(\mu\pm\sqrt{|\delta|^{2}+\mu^{2}}\bigr)^{2}}} \\
      0 \\
      0 \\
      0
   \end{array}
   \right),
\nonumber \\ 
   U^{\pm}_{5}
&=
   \left(
   \begin{array}{c}
      0 \\
      0 \\
      0 \\
      0 \\ 
      \hline
      0 \\
      0 \\
      0 \\
      \dfrac{-i\delta\bigl(\mu\pm\sqrt{|\delta|^{2}+\mu^{2}}\bigr)}
      {|\delta|\sqrt{|\delta|^{2}+\bigl(\mu\pm\sqrt{|\delta|^{2}+\mu^{2}}\bigr)^{2}}} \\ 
      \hline
      0 \\
      0 \\
      0 \\
      \dfrac{|\delta|^{2}}{|\delta|\sqrt{|\delta|^{2}+\bigl(\mu\pm\sqrt{|\delta|^{2}+\mu^{2}}\bigr)^{2}}} \\ 
      \hline
      0 \\
      0 \\
      0 \\
      0
   \end{array}
   \right),
\quad 
   U^{\pm}_{6}
=
   \left(
   \begin{array}{c}
      0 \\
      0 \\
      0 \\
      0 \\ 
      \hline
      0 \\
      0 \\
      \dfrac{-i\delta\bigl(-\mu\pm\sqrt{|\delta|^{2}+\mu^{2}}\bigr)}
      {|\delta|\sqrt{|\delta|^{2}+\bigl(-\mu\pm\sqrt{|\delta|^{2}+\mu^{2}}\bigr)^{2}}} \\ 
      0 \\ 
      \hline
      0 \\
      0 \\
      \dfrac{|\delta|^{2}}{|\delta|\sqrt{|\delta|^{2}+\bigl(-\mu\pm\sqrt{|\delta|^{2}+\mu^{2}}\bigr)^{2}}} \\
      0 \\ 
      \hline
      0 \\
      0 \\
      0 \\
      0
   \end{array}
   \right),
\nonumber \\ 
   U^{\pm}_{7}
&=
   \left(
   \begin{array}{c}
      0 \\
      0 \\
      0 \\
      0 \\ 
      \hline
      0 \\
      \dfrac{-i\delta\bigl(-\mu\pm\sqrt{|\delta|^{2}+\mu^{2}}\bigr)}
      {|\delta|\sqrt{|\delta|^{2}+\bigl(-\mu\pm\sqrt{|\delta|^{2}+\mu^{2}}\bigr)^{2}}} \\ 
      0 \\
      0 \\ 
      \hline
      0 \\
      \dfrac{|\delta|^{2}}{|\delta|\sqrt{|\delta|^{2}+\bigl(-\mu\pm\sqrt{|\delta|^{2}+\mu^{2}}\bigr)^{2}}} \\
      0 \\
      0 \\ 
      \hline
      0 \\
      0 \\
      0 \\
      0
   \end{array}
   \right),
\quad 
   U^{\pm}_{8}
=
   \left(
   \begin{array}{c}
      0 \\
      0 \\
      0 \\
      0 \\ 
      \hline
      \dfrac{-i\delta\bigl(\mu\pm\sqrt{|\delta|^{2}+\mu^{2}}\bigr)}
      {|\delta|\sqrt{|\delta|^{2}+\bigl(\mu\pm\sqrt{|\delta|^{2}+\mu^{2}}\bigr)^{2}}} \\ 
      0 \\
      0 \\
      0 \\ 
      \hline
      \dfrac{|\delta|^{2}}{|\delta|\sqrt{|\delta|^{2}+\bigl(\mu\pm\sqrt{|\delta|^{2}+\mu^{2}}\bigr)^{2}}} \\
      0 \\
      0 \\
      0 \\ 
      \hline
      0 \\
      0 \\
      0 \\
      0
   \end{array}
   \right).
\label{eq:SU2PF_hamiltonian_ansatz_Dirac_point_wf}
\end{align}
Defining the chiral operator in the Nambu-Gor'kov space by
\begin{align}
   \Gamma_{5}
=
   \left(
   \begin{array}{cc}
      \gamma_{5} \mathbbm{1} & 0 \\
      0 & \gamma_{5} \mathbbm{1}
   \end{array}
   \right),
\end{align}
with the $2\times2$ identity operator $\mathbbm{1}$ for $(\psi_{1},\psi_{2})$ in the emergent chiral space,
we find that $U^{^{\pm}}_{1,2}$ and $U^{\pm}_{5,6}$ have a negative parity and $U^{\pm}_{3,4}$ and $U^{\pm}_{7,8}$ have a positive parity, as shown by
\begin{align}
   \Gamma_{5} U^{\pm}_{1} = - U^{\pm}_{1}, \quad 
   \Gamma_{5} U^{\pm}_{2} = - U^{\pm}_{2}, \quad 
   \Gamma_{5} U^{\pm}_{3} = U^{\pm}_{3}, \quad 
   \Gamma_{5} U^{\pm}_{4} = U^{\pm}_{4}, \nonumber \\ 
   \Gamma_{5} U^{\pm}_{5} = - U^{\pm}_{5}, \quad 
   \Gamma_{5} U^{\pm}_{6} = - U^{\pm}_{6}, \quad 
   \Gamma_{5} U^{\pm}_{7} = U^{\pm}_{7}, \quad 
   \Gamma_{5} U^{\pm}_{8} = U^{\pm}_{8}.
\end{align}

\bibliography{chiral_superfluid_neutron_star,refs}

\end{document}